%% file: MAIN.tex
\documentclass[12pt]{article}

\usepackage{graphicx}   
\usepackage{subcaption} 
\usepackage{graphicx}   
\usepackage{makecell}   
\usepackage{booktabs}   
\usepackage{xcolor}     
\usepackage{pifont}     
\usepackage{array}      
\usepackage{enumitem}   
\usepackage{longtable}  
\usepackage{rotating}   
\usepackage{placeins}   
\usepackage{float}      
\usepackage{afterpage}  
\usepackage{rotfloat}   
\usepackage{tikz}       
\usepackage{authblk}    
\usepackage[american]{babel}
\usepackage{hyperref}
\usepackage{amsmath}
\usepackage{array}
\usepackage{caption} 
\usepackage{tabularx}
\usepackage{natbib}
\usepackage[a4paper, margin=1in]{geometry}
\usepackage{multirow}
\usepackage{multicol}
\usepackage{amssymb}
\usetikzlibrary{calc}
\usetikzlibrary{positioning}

\newcommand{\erin}[1]{\textcolor{black}{{#1}}}

\newcommand{\nk}[1]{\textcolor{black}{{#1}}}
\newcommand{\emily}[1]{\textcolor{black}{{#1}}}
\definecolor{red}{RGB}{255,0,0} 
\definecolor{blue}{RGB}{0,0,255} 
\def\checkmark{\tikz\fill[red, scale=0.4](0,.35) -- (.25,0) -- (1,.7) -- (.25,.15) -- cycle;}
\newcommand{\bluecheckmark}{\tikz\fill[blue, scale=0.4](0,.35) -- (.25,0) -- (1,.7) -- (.25,.15) -- cycle;}

\title{PADTHAI-MM: Principles-based Approach for Designing Trustworthy, Human-centered AI using MAST Methodology}

\author[1,2]{\normalsize Myke C. Cohen}

\affil[1]{\normalsize Human Systems Engineering, Arizona State University, AZ, USA}

\affil[2]{Aptima, Inc., MA, USA}

\author[3]{Nayoung Kim}

\affil[3]{School of Computing and Augmented Intelligence, Arizona State University, AZ, USA}

\author[3]{Yang Ba}

\author[1]{Anna Pan}

\author[1]{Shawaiz Bhatti}

\author[1]{Pouria Salehi}

\author[4]{James Sung}

\affil[4]{U.S. Department of Homeland Security, VA, USA}

\author[5]{Erik Blasch}

\affil[5]{Air Force Office of Scientific Research, VA, USA}

\author[6]{Mickey V. Mancenido}

\affil[6]{School of Mathematical and Natural Sciences, Arizona State University, AZ, USA}

\author[1]{Erin K. Chiou}

\begin{document}



\date{} 

\maketitle

\input{sections/0_Abstract}
\input{sections/1_Introduction}
\input{sections/2_RelatedWorks}
\input{sections/3_Framework}
\input{sections/4.1_CaseStudy}
\input{sections/4.2_Validation}
\input{sections/5_Discussion}
\input{sections/6_Conclusion}
\input{sections/7_Acknowledgments}

\section{Disclosure of Potential Conflicts of Interest}
The authors have no conflicts of interest to declare.

\bibliographystyle{apalike2}
\bibliography{refs}
\input{sections/appendixA}

\input{sections/appendixB}

\input{sections/appendixC}
\input{sections/appendixE}
\input{sections/appendixD}

\end{document}

%% file: sections/0_Abstract.tex
Despite an extensive body of literature on trust in technology, designing trustworthy AI systems for high-stakes decision domains remains a significant challenge, further compounded by the lack of actionable design and evaluation tools. The Multisource AI Scorecard Table (MAST) was designed to bridge this gap by offering a systematic, tradecraft-centered approach to evaluating AI-enabled decision support systems. Expanding on MAST, we introduce an iterative design framework called \textit{Principles-based Approach for Designing Trustworthy, Human-centered AI using MAST Methodology} (PADTHAI-MM). We demonstrate this framework in our development of the Reporting Assistant for Defense and Intelligence Tasks (READIT), a research platform that leverages data visualizations and natural language processing-based text analysis, emulating an AI-enabled system supporting intelligence reporting work. To empirically assess the efficacy of MAST on trust in AI, we developed two distinct iterations of READIT for comparison: a High-MAST version, which incorporates AI contextual information and explanations, and a Low-MAST version, akin to a ``black box'' system. This iterative design process, guided by stakeholder feedback and contemporary AI architectures, culminated in a prototype that was evaluated through its use in an intelligence reporting task. We further discuss the potential benefits of employing the MAST-inspired design framework to address context-specific needs. We also explore the relationship between stakeholder evaluators' MAST ratings and three categories of information known to impact trust: \textit{process}, \textit{purpose}, and \textit{performance}. Overall, our study supports the practical benefits and theoretical validity for PADTHAI-MM as a viable method for designing trustable, context-specific AI systems.

%% file: sections/1_Introduction.tex
\section{Introduction}

Rapid advances in artificial intelligence (AI) have broadened the use of decision support systems (DSS) in high-criticality fields such as healthcare, national security, and finance. In these domains, AI-enabled decision support systems (AI-DSSs) offer significant potential to enhance decision-making effectiveness and system efficiency. However, the low tolerance for errors in these areas highlights the need to balance users' perceptions of AI's benefits with the risks of system failures. This emphasizes the increasingly important role of trust in AI when designing effective AI-DSSs.

Recent work have shown a direct association between human trust in AI and the system's transparency~\citep{21,22,23}, i.e. the AI's ability to clearly explain its processes, decisions, and limitations~\citep{millerTrustTransparencyExplanation2021}. However, state-of-the-art AI-DSSs are often designed using ``black-box'' deep learning models, that enable complex functionalities but are generally not human-interpretable~\citep{duran2021afraid, london2019artificial}. To address transparency and trustworthiness, researchers have developed methods to generate algorithmic interpretations of black-box model outputs without compromising the system's core capabilities~\citep{laiHumanPredictionsExplanations2019}. Some critics argue that interpretations alone are insufficient for critically examining AI outputs, especially with respect to human values and ethical considerations~\citep{shenTrustAIInterpretability2022}. Theoretical frameworks have been proposed to incorporate trust-related considerations in the design of reliable and responsible AI~\citep{lee2004trust, malleChapterMultidimensionalConception2021}. Furthermore, recent scholarship has increasingly emphasized that these design considerations should be context-specific, human-centered, and must align with organizational missions and goals (\citealp{cech2021agency,kowald2024,17,12}; cf. \citealp{munnUselessnessAIEthics2023}).

A limitation of current principles-based design frameworks lies in their reliance on abstract concepts, which makes it challenging to translate into concrete and actionable system requirements~\citep{stanley2024}. Additionally, the need to account for context and mission-specific criteria of what makes an AI system trustworthy renders some design frameworks impractical for guiding actual design processes~\citep{hagendorff2020} because they were written to be broadly generalizable rather than for specific domains~\citep{prem2023}. 

Tradecraft standards are ubiquitous in well-developed, high-stakes work domains, outlining operating procedures, principles, professional benchmarks, and best practices to guide practitioners in delivering effective, compliant, and high-quality outcomes. Examples include Intelligence Community Directives (ICD), the Federal Aviation Administration's Airworthiness Directives and Checklists, and the World Health Organization's Clinical Practice Guidelines. When linked to concrete processes, tools, evaluation metrics, and stakeholder roles, these materials can guide AI-DSS designers to address design and evaluation considerations for trustworthy AI-DSSs~\citep{yildirimInvestigatingHowPractitioners2023a, deng2023investigating}. However, tradecraft-derived guidelines often fail to explicitly connect design recommendations with empirical and theoretical research, creating gaps between the AI transparency strategies employed by practitioners and research-oriented industry standards \citep{schorMindGapDesigners2024}. To close these gaps, researchers and practitioners must be able to adapt tradecraft principles into concrete guidelines that not only incorporate state-of-the-art algorithmic solutions but also explicitly align design alternatives with users' theoretical trust-related information needs.


In this paper, we document our formulation and validation of an iterative design framework for intelligence AI-DSS use cases, called the ``\textbf{P}rinciples-based \textbf{A}pproach for \textbf{D}esigning \textbf{T}rustworthy, \textbf{H}uman-centered \textbf{AI} using \textbf{M}AST \textbf{M}ethodology'' (PADTHAI-MM). PADTHAI-MM was derived from the Multisource AI Scorecard Table (MAST; \citealp{blasch2021multisource}), an AI trustworthiness evaluation tool based on the intelligence community's tradecraft analytic standards. Using PADTHAI-MM, we developed and evaluated two versions of a text summarization system called the Reporting Assistant for Defense and Intelligence Tasks (READIT). As MAST embodies the intelligence community's domain-specific definition of what makes AI trustworthy, we used its criteria and sub-criteria to operationalize design requirements for the particular application. To illustrate PADTHAI-MM's validity as a design framework, a ``High-MAST'' version was designed to meet the Intelligence Community's trustworthiness criteria, while the ``Low-MAST'' version was designed as a minimum viable product that only contained functionalities for task completion. We posit that independent users (both experts and non-experts uninvolved in the design process) who interact with the ``High-MAST'' version will provide higher ratings based on the MAST criteria, clearly distinguishing it from the lower ratings expected for the ``Low-MAST'' version. Such findings would indicate that the proposed design framework effectively and reliably translates the underlying MAST principles into actionable, evaluable design outcomes aligned with the intelligence community's definition of AI trustworthiness. Finally, to establish construct and convergent validity of the design process, we compare resulting MAST ratings of system trustworthiness with user perceptions of information categories known to influence user trust in automated systems.



%% file: sections/2_RelatedWorks.tex
\section{Related Works}

\subsection{\textbf{Trust, Transparency, Interpretability, \& Explainable AI}}
Trust has been defined as the attitude that another agent can help one achieve their goals in a risky situation \citep{lee2004trust, mayer1995integrative}. Decades of empirical research suggest that trust is related to various decision-making biases with imperfect automation \citep{chiou2023trusting, lee2004trust, leeTrustSelfconfidenceOperators1994, mcguirlSupportingTrustCalibration2006, parasuraman1997humans, riegelsbergerMechanicsTrustFramework2005, tenhundfeldAssessmentTrustAutomation2022, 8, devisserAutomationAutonomyImportance2018}. For example, people tend to trust a DSS's recommendations despite the presence of external indicators of an error \citep{lyellAutomationBiasVerification2017, cummingsAutomationBiasIntelligent2015a}. However, once aware that a DSS has made a mistake, people tend to immediately distrust it \citep{dietvorstAlgorithmAversionPeople2015, kahrTrustRecoveryJourney2024}. Because of such human difficulties in gauging when to rely on DSSs, trustworthiness is universally considered an important design objective (\citealp{8}; cf. \citealp{boltonTrustNotVirtue2024}).

Access to trust-relevant information is a precursor to developing well-calibrated trust \citep{lee2004trust, millerTrustTransparencyExplanation2021, 21, 22}. However, there is a trade-off between AI system transparency and decision-making performance (\citealp{castelvecchiCanWeOpen2016, guptaIntelligentDataAnalysis2020}; cf. \citealp{rudin2019stop, 8882211}). As such, black-box AI architectures are the default models in AI-DSSs, making it challenging for designers to consider an operator's information requirements with respect to understanding the underlying logic behind system recommendations and their overall reliability~\citep{london2019artificial, 23}. 

Efforts to improve the trustworthiness of black-box models have led to design strategies that augment displayed outputs with additional information. The most prominent of these are Explainable Artificial Intelligence (``XAI''; ~\citealp{Gunning_Aha_2019}) methods, which overlay black-box models with post-hoc explanations to make their outputs human-comprehensible. Such explanations can range from representative predictions to IF-THEN rules or feature contribution analyses~\citep{1,2,3}. Scholarship on AI transparency continues to prioritize methods for interpreting black-box models over advancing ``white-box'' alternatives with traceable decision logic~\citep{hassijaInterpretingBlackBoxModels2024, saranyaSystematicReviewExplainable2023}. Yet, the effectiveness of XAI and similar approaches have yielded mixed empirical results for improving the trustability of AI-DSSs, in part because they address only one aspect that can affect trust~\citep{weberXAITrouble2024}. 

Many studies have shown that algorithmic interpretations of AI outputs can make users perceive a system as more trustworthy without substantially improving decision-making performance \citep{6, laiHumanPredictionsExplanations2019, zhangEffectConfidenceExplanation2020}. There now exist many types of XAI design strategies \citep{wangEffectsExplanationsAIAssisted2022}, but it remains unclear which of these most effectively inform trust, or how to determine their appropriateness for different contexts and user needs. Some have argued that excessive attention has been devoted to model interpretability, neglecting the role of interoperability in trust-centered design objectives~\citep{rudin2019stop, shenTrustAIInterpretability2022}. For one, there is an abundance of model trustworthiness evaluation techniques that rely on using purely algorithmic measures (e.g., \citealp{chengThereHopeAll2020, chengGeneralTrustFramework2021, 10.1145/3236009}) or compare a model's explanations to those generated by benchmark algorithms like LIME, SHAP, TabNet, and Anchors (\citealp{mirzaeiExplainableAIEvaluation2024}). Such techniques rarely incorporate input from end users. Consequently, this paper's objective is to bridge the gap between black-box AI interpretation methods and other critical aspects of human-AI interaction that influence trust.

\subsection{\textbf{Context and Interaction Design Considerations}}


Seminal research on human-automation interaction~\citep{lee2004trust} proffers that  trust in an AI-DSS depends on people's understanding of the technology's \textit{processes} (i.e., how it performs tasks), \textit{purpose} (i.e., why the system was developed), and \textit{performance} (i.e., how effectively it carries out those tasks). Recent work has indicated that these three information categories can inform technology designers and evaluators about how operators trust a system and make decisions about its use. Much of this work has been applied to develop various measures of trust in DSSs (e.g., \citealp{chancey2017trust, meyerTrustRelianceCompliance2013}) that designers can use to improve system features between use periods or design iterations. Real-time, quantitative trust models can theoretically be built into AI-DSSs that can then adapt information displays to better support appropriate trust levels \citep{devisserMutuallyAdaptiveTrust2023}. However, many existing examples (e.g., \citealp{cavorsiExploitingTrustResilient2023, cheng2022trustworthiness}) lack a direct link to empirically-observed human trust phenomena, limiting their practical applicability and ecological validity. 

Users tend to develop their understanding of system trustworthiness through repeated interactions \citep{chiou2023trusting}. Thus, in addition to making AI processes interpretable, designing for trustworthiness also requires considering how the interactive features of AI-DSSs influence the formation of trust \citep{de2020towards}. \cite{8} suggests that trust-building human-AI interactions should be designed after end-user values and expectations. However, system developer teams do not always have the skills, tools, or resources to adapt end-user inputs into specific design criteria \citep{millerExplanationArtificialIntelligence2019}. AI-DSS developers are instead increasingly reliant on general design guidelines that outline users' trust information needs \citep{deng2023investigating}. Indeed, \cite{yildirimInvestigatingHowPractitioners2023a} found that trust and explainability are the foremost topics for which AI designers reference Google's "People + AI Research" (PAIR) Guidebook \citep{googlepairPeopleAIGuidebook2021}. However, they also found that its lack of domain-specificity leads practitioners to create ad hoc benchmarks for system trustworthiness.

In high-stakes domains, trust has become a design dilemma amid increasing objections to the use of black-box AI architectures \citep{23, rudin2019stop} and the emergence of legal regulations surrounding AI transparency, such as the EU AI Act. On the one hand, black-box models have been consistently shown to mitigate human errors in high-stakes decision tasks. These include tendencies to misjudge risk likelihoods \citep{stevensonAlgorithmicRiskAssessment2024} and detect deceptive manipulations in digital media \citep{laiHumanPredictionsExplanations2019, naberComparingHumanAnalytic2024}. On the other hand, employing inscrutable AI-DSS architectures to aid in such tasks introduces the risks of over reliance \citep{coieraTechnologyCognitionError2015}. 
Risk tolerance is one of many system design considerations that are known to vary between high-stakes domains \citep{hendersonRiskTakingProfessionalGroups2021}; nonetheless, recent design guidelines for trustworthy DSSs (e.g., \citealp{stoneDesignThinkingFramework2022}) do not account for these differences. Domain-specific tradecraft materials may outline risk tolerance levels and related qualities, but are generally an underexplored resource for informing technology design methods.

\subsection{\textbf{Applying Tradecraft Principles: The MAST Criteria}}
The U.S. Intelligence Community, under the purview of the U.S. Department of Homeland Security's Office of the Director of National Intelligence (ODNI), is an example of a high-stakes domain where tradecraft standards have borne design guidance for AI trustworthiness. \cite{odniIntelligenceCommunityDirective2015} specifies analytic tradecraft standards in Intelligence Community Directive (ICD) 203, that include transparency qualities commonly used in XAI approaches, such as Uncertainty and Visualization, as well as user-oriented qualities like Customer Relevance. These standards formed the basis for the Multisource AI Scorecard Table (MAST; \citealp{blasch2021multisource})---a nine-item, four-level scale for assessing AI trustworthiness (0-3, with 0 representing “poor” and 3 “excellent”)---devised by a group from the 2019 Public-Private Analytic Exchange program sponsored by the Department of Homeland Security’s Office of Intelligence and Analysis, on behalf of ODNI. 

MAST is one of few domain-specific, tradecraft-derived measures for AI trustworthiness that have been validated against existing theoretically-derived and empirically-tested trust questionnaires \citep{chiou22}. In \cite{salehiTrustworthyAIEnabledDecision2024}, we showed that MAST-inspired transparency designs can improve operator perceptions of system trustworthiness despite not guaranteeing improved overall system outcomes, similar to XAI-based system designs. Closer analyses in a face matching use case showed that aspects of human-AI interaction help explain when and how MAST-inspired system features contribute to improved decision outcomes \citep{cohenMultiMeasureTrustCalibration2024}. However, there exists no detailed framework for translating MAST and other principles-based tools into actionable guidelines for balancing algorithmic transparency, human trust phenomenology, and domain-specific requirements. In this paper, we address this gap by:
\begin{enumerate}
    \item  Translating theoretical and MAST-derived system trustworthiness principles into actionable design guidelines;
    \item Demonstrating the use of MAST to design and evaluate domain-specific features; and
    \item Validating how a MAST-inspired design process can align design iterations with stakeholder trust considerations.
\end{enumerate}


%% file: sections/3_Framework.tex
\section{PADTHAI-MM Framework}
The Principles-based Approach for Designing Trustworthy, Human-centered AI using MAST Methodology (PADTHAI-MM) is a design framework that integrates system developers' AI and data knowledge with trust scholarship to develop trustworthy AI-DSSs for intelligence use cases, following AI trustworthiness principles outlined in the MAST criteria \citep{blasch2021multisource}. PADTHAI-MM comprises nine major steps and borrows from iterative design frameworks for product development~\citep{ulrichProductDesignDevelopment2020}. We posit that each step contributes to improving system efficacy and alignment with organizational values, with design ideation and execution steps for connecting these qualities to corresponding AI features.

\begin{figure*}[t!]
    \centering
    \includegraphics[width=\textwidth]{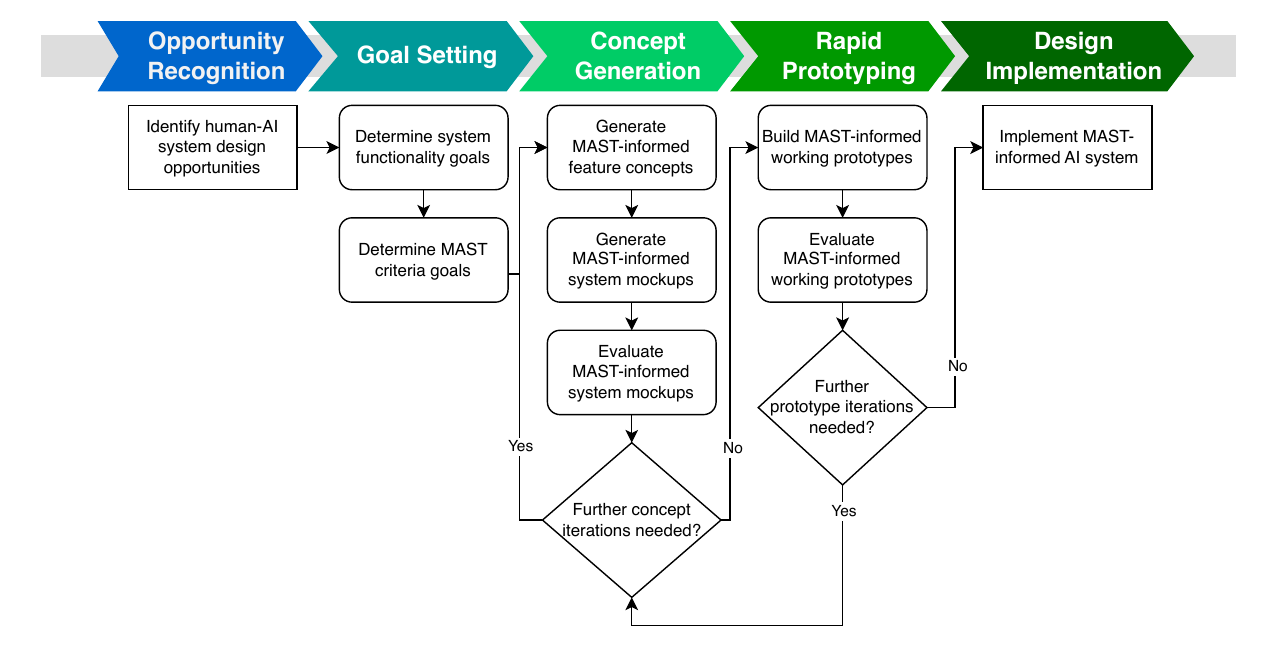}
    \caption{PADTHAI-MM Design Framework}
    \label{fig:3component}
\end{figure*}

\autoref{fig:3component} summarizes the PADTHAI-MM framework.
First is \textit{Opportunity Recognition}, through which AI-DSS use cases and contexts are identified alongside the needs and expectations of operators and other peripheral stakeholders where the AI-DSS will be deployed. This is to ensure that designs developed in subsequent steps address the MAST criterion of Customer Relevance. Afterwards, \textit{Goal Setting} steps follow to aid design teams identify system functionality goals and desired trustworthiness levels through setting target MAST ratings, guided by integrating stakeholder inputs, system evaluations, and design teams' knowledge of feasible AI solutions. Once system goals are specified, \textit{Concept Generation} steps will iteratively generate and select individual features and full system concepts. Selected designs will then be used in \textit{Rapid Prototyping} steps, incorporating stakeholder inputs in the design and evaluation of working system prototypes. Finally, \textit{Design Implementation} takes place after stakeholder evaluations confirm that one or more system prototypes now satisfy system functionality and MAST goals. The following sections describe each of these steps in more detail.



\paragraph*{\textbf{Step 0: Opportunity Recognition}}
A precursor step is to select viable opportunities for implementing an AI-DSS to aid in the overall objectives of a work system. This involves identifying and systematically engaging with key stakeholders within the work system through interviews, focus group discussions, and periodic surveys (see \citealp{iandoloStakeholderEngagementManaging2024} for a methodological example). This process aims to collect detailed insights regarding current workflows when they exist, highlighting specific challenges and inefficiencies within the system. Note that this involves considering the existing AI-DSS implementations within a system, which may entail evaluating personnel and machine capabilities.

\paragraph*{\textbf{Step 1: Set system functionality goals}}
Step 1 establishes a concrete set of tasks to design the AI-DSS. This entails two components: (1) identifying which work processes could be addressed or enhanced by an AI-DSS, and (2) assessing how the integration of AI-DSS might introduce, expand, enhance, or eliminate certain human work processes. 

Identifying candidate processes involves breaking down the work system into a set of tasks. Formal techniques include operational sequence diagramming \citep{kurkeOperationalSequenceDiagrams1961} and hierarchical task analysis \citep{stantonHierarchicalTaskAnalysis2006}. To help assess the utility of AI-DSS solutions, design teams must define parameters for acceptable system outputs and reliability expectations (e.g., permissible error rates within specified operational periods) and assess the potential implications of suboptimal outputs or system failures. Moreover, teams must also consider ethical, regulatory, and legal constraints that delimit the system's operational boundaries. Performance-based and regulatory considerations tend to be outlined in industry standards; in niche use cases, stakeholder inputs may serve as a quick reference.

Once the work system is broken down into a set of tasks with defined metrics, design teams can determine AI-DSS system functionality goals by considering potential implementation. For existing systems, the determination process can be based on available historical data; if none exist, time studies and work sampling methods \citep{matiasWorkMeasurementPrinciples2001} can generate low-cost benchmarks. For novel systems, this may involve exploring various task assignment configurations between operators and their prospective AI-DSS counterparts, consulting subject matter experts, or using formal techniques (e.g., failure mode and effect analyses; \citealp{stamatisFailureModeEffect2003}) if within design team's capabilities. Finally, documenting high-level system flowcharts is encouraged to aid in subsequent design steps.

\paragraph*{\textbf{Step 2: Determine MAST criteria goals}} 
Step 2 establishes trustworthiness-specific design goals using MAST. This involves evaluating the relevance of each MAST criterion to each system objectives identified in Step 1, followed by establishing MAST rating targets for the system.   \emily{Multiple Decision Criteria Analysis (MCDA; \citealp{ferreiraMeasuringTradeoffsCriteria2013}) frameworks can facilitate this evaluation process by providing a structured approach.} Understanding the relative importance between criteria is crucial for navigating potential design trade-offs in subsequent design phases.

We emphasize that this step entails specifying desired ratings for each MAST criterion and an overall MAST score. Target MAST ratings establish benchmark levels for assessing and iterating upon system design concepts in subsequent steps. In cases where the process involves redesigning an existing AI-DSS, evaluate the current system using MAST to pinpoint areas for improvement. 

\paragraph*{\textbf{Step 3: Generate MAST-informed feature concepts}}
Step 3 creates an initial set of system feature concepts to address individual system functionality and trustworthiness goals. For example, data summarization functionalities may be conceptually addressed through on-demand statistical outputs; to address MAST target ratings, additional features to consider may include visualized displays or textual reports. To create a sufficiently broad range of options for subsequent design steps, design teams should couple systematic exploration of external sources (e.g., benchmarking similar DSSs, searching patents and literature, consulting lead users and experts) and internal brainstorming for possible new approaches. We refer the reader to \citet[ch.7]{ulrichProductDesignDevelopment2020} for a suggested step-by-step procedure to facilitate this process, which includes recommended ways to address potential group decision-making biases or conflicts.

As feature concepts are generated for each candidate process identified in Step 1, their MAST ratings may also be estimated to determine their potential contributions to the MAST goals identified in Step 2. \emily{Teams engaged in developing critical, industrial-scale systems would benefit from documenting these preliminary ratings to allow for a more rigorous evaluation of the potential variation range for each MAST criterion, and to possibly refine or reconsider the features to reduce the range. This would comprise an additional iterative step in the design process, although not explicitly depicted in \autoref{fig:3component}.} Features may address multiple candidate processes or multiple MAST criteria; thus, design teams should map how feature concepts address each of its target candidate processes and assign MAST rating estimates. Note that feature concepts generated at this step, including their interoperability, will be further refined in subsequent iterations before their inclusion in system-level configurations. 

\paragraph*{\textbf{Step 4: Generate MAST-informed system mockups}}
Step 4 creates full system mockups by combining individual feature concepts from Step 3. In doing so, design teams must consider the full suite of system functionality goals from Step 1 and their corresponding feature concepts. First, design teams will generate feature combinations based on criteria such as:
\begin{itemize}
    \item \textit{Feature inter-compatibility}, including their potential to confuse users in assessing system trustworthiness, usability, or algorithmic performance. Tools like Hauser and Clausing's (\citeyear{hauser66house}) ``House of Quality'' are recommended to streamline assessments of feature inter-compatibility.
    \item \textit{Feature workflows and interaction modalities} and their performance impacts, especially regarding process or information redundancies.
    \item \textit{Feasibility} of displaying features simultaneously, considering interface constraints and human perceptual limits \citep{Fitts_1954}.
\end{itemize}

Feasible feature combinations are designated as \textit{full system concepts} (as opposed to feature concepts from Step 3). For each system concept, design teams will generate preliminary mockups (e.g., sketches, storyboards) to illustrate how users may interact with the system, noting required inputs and anticipated outputs. Design teams must then consolidate, eliminate, and regenerate these system concepts until a manageable number of alternatives is reached. To guide this process, structured decision-making techniques like concept scoring or axiomatic matrices (cf. \citealp{xiaoComparisonConceptSelection2007}) are recommended.

Finally, audiovisual or storyboard-like mockup presentations of the selected system concepts should be prepared to facilitate stakeholder evaluations in later steps. These presentations will be essential for collecting stakeholder feedback on how each configuration aligns with the system functionality and objectives established in Steps 1 and 2. Importantly, mockup presentations should emphasize end-user workflows through sequentially-accurate representations of realistic scenarios and avoid explicitly communicating specific design goals, rationales, intended process-feature pairings, and system trustworthiness levels. This approaches ensures that stakeholders can focus on evaluating the practical use and functionality of each system concept from their perspectives with minimal design team influence.

\paragraph*{\textbf{Step 5: Evaluate MAST-informed system mockups}}
Step 5 establishes assessments of full system mockups relative to system functionality and MAST rating goals set in Steps 1 and 2. This step involves recruiting participants outside the design team who will review the mockup presentation materials generated in Step 4 and provide evaluative feedback through structured or semi-structured study protocols. Ideally, design teams should enlist stakeholders (e.g., operators, training personnel, subject matter experts) as evaluative study participants. However, preliminary data collected from general population participants can still provide useful insights.

Data collected \textit{must} include participants' MAST ratings of the system mockup, as well as information related to its perceived utility within its intended use cases. Interviews and focus group discussion formats can also yield qualitative data and uncover unforeseen trade-offs or design considerations. 


\paragraph*{\textbf{Step 6: Iterate MAST-informed feature \& system concepts}}
Step 6 determines subsequent steps based on stakeholder feedback on system concept mockups generated in Step 4. Design teams should analyze the data collected in Step 5 and determine whether desired system functionality and MAST rating goals have been met from end-user perspectives. The results of the analysis should be compared to the system functionality and MAST criteria goals determined in Steps 1 and 2, respectively, guided by the mapping of system features to specific MAST criteria mapping in Step 3. 

Depending on the analysis results, the design team (ideally including AI engineers, human systems engineers, and task domain experts) should recommend which system alternatives are ready for prototyping, need further design iterations, or should be discarded. System concepts whose mockups were rated as meeting system functionality goals but not MAST criteria ones should be considered as needing further design iterations, roughly following the procedure outlined in Steps 3 and 4. If results show no system concept alternatives meet either set of design goals, however, the design team is strongly encouraged to restart from Step 1 and reconsider which AI-DSS opportunities and system functionalities to prioritize. Such instances may be rare, but are illustrated in our case application of PADTHAI-MM. We also note a need for nuanced interpretation of data gathered in Step 5, as it only involves participants evaluating system mockups, not working prototypes. Design teams must account for participant population characteristics when interpreting results and determining next design steps---ideally, with direct involvement of stakeholders.



\paragraph*{\textbf{Step 7: Build MAST-informed working prototypes}}
Step 7 develops into working prototypes the system concepts that were determined to meet system functionality and MAST rating goals at the end of Step 6. At this stage, core system features will be implemented by the design team. The goal is to create preliminary working AI-DSSs that users can interact with to perform core system functionalities, with access to features that supply trust-related information. 

We caution design teams against overdesigning working prototypes, as multiple rounds of rapid prototyping may be needed before a system design is ready for implementation. Minor differences between mockup versions and prototypes should also be expected as unforeseen constraints are discovered in the process. However, these should be kept to a minimum and documented by the design team as a reference for interpreting user evaluations of working prototypes, outlined in Step 8. 

\paragraph*{\textbf{Step 8: Evaluate MAST-informed working prototypes}}
Step 8 establishes assessments of working prototypes developed in Step 7 relative to system functionality and MAST rating goals set in Steps 1 and 2. The primary objective of this step is to observe how prospective system operators interact with AI-DSS prototypes to conduct intended tasks, including how useful and trustworthy they perceive the system to be. Design teams will design and conduct evaluative studies, recruiting participants who will complete one or more decision-making workflows with an AI-DSS prototype. Task scenarios should be designed to mirror the operational range of decisions and environmental conditions. As in Step 5, data collected should include, but is not limited to, participants' MAST ratings of the system, as well as responses to questionnaires on system usability and perceived utility. Participants should also be recruited from stakeholder populations to ensure that evaluations approximate those from likely end-users.

Design teams are also highly encouraged to collect measures of the extent to which AI-DSS prototypes influence and improve operators' decision-making. This involves translating the system functionality goals and metrics established in Step 1 into specific metrics that can be gathered through the working prototypes. Observation-based measures of decision adoption (e.g., reliance or compliance with AI recommendations \citealp{meyerTrustRelianceCompliance2013}) and effectiveness (e.g., accuracy, quality, throughput) are suggested.

\paragraph*{\textbf{Step 9: Determine iteration and implementation needs}}
Step 9 determines whether further design iterations are needed before full system implementation. Upon evaluating the working prototypes in Step 8, the design team should conduct statistical analyses of questionnaire responses and performance metrics, comparing these findings against the system functionality and MAST rating goals established in Steps 1 and 2. As with Step 6, we advise design teams to interpret analyses of data collected in Step 8 with caution. There are known risks that users may conflate early prototypes with finished products, rating them unfavorably when they only deliver core functionalities; conversely, the same prototypes may also function more effectively than final AI-DSSs \citep[ch. 9]{ulrichProductDesignDevelopment2020}.

Should one or more prototypes meet system performance and MAST rating benchmarks, the design team and stakeholders must determine a plan for finalizing and implementing the selected prototype(s). We refer the reader to \citet[ch. 24]{booherHandbookHumanSystems2003} for an overview of considerations in implementing new AI-DSS solutions, which is beyond the scope of PADTHAI-MM. 

Conversely, if none of the prototypes satisfy the performance or MAST criteria goals, design teams should refine existing prototypes, roughly following Step 7, and conduct another round of user testing as outlined in Step 8. If repeated prototyping attempts fail to yield satisfactory results, the design team may collectively conclude that other system concepts should be considered. System concepts previously discarded in Steps 4 or 5 may be rapidly built into working prototypes and tested with users as a first step. However, if discarded system concepts are too similar to unsatisfactory prototypes, design teams may consider altering the suite of system features, processes, and goals. Such cases may indicate insufficient stakeholder involvement in Steps 1 and 2, or a lack of viewpoint diversity within the design team.

%% file: sections/4.1_CaseStudy.tex
\section{Framework Validation}
In this section, we illustrate an application of the PADTHAI-MM framework in our development of a research platform and AI-DSS use case for intelligence reporting, called \textbf{READIT} (\textbf{RE}porting \textbf{A}ssistant for \textbf{D}efense and \textbf{I}ntelligence \textbf{T}asks). Based on the final outcome of this design case study, we then demonstrate how PADTHAI-MM helps to address users' trust information needs by relating the resulting system features to process, purpose, and performance information, as delineated by \cite{lee2004trust}.

\subsection{Design Case Study: \textit{READIT}}

\paragraph*{\textbf{Step 0: READIT Opportunity Recognition}}
We began by considering how AI-DSSs can aid intelligence analysts in identifying potential risks to national security by monitoring social media platforms. This use case was selected upon consultation with project stakeholds and subject matter experts from the intelligence community. Another reason for selecting a natural language processing (NLP) use case was to maximize ecological validity, as MAST was originally developed for NLP analytic applications.


\paragraph*{\textbf{Step 1: Setting READIT 1.0 system functionality goals}}
To determine system functionality goals, our team identified likely key processes involved in detecting security threats through monitoring public sentiment analysis in social media. Based on this process, our initial attempt (``READIT 1.0'') aimed toward three primary functions: (1) rapidly curating a list of relevant social media content based on user-supplied keywords; (2) determining the influence of each curated content, based on engagement metrics such as post reactions and reposts; and (3) displaying an AI-generated extractive summary that encapsulates the essence of the most relevant and influential posts.


\paragraph*{\textbf{Step 2: Determining READIT 1.0 MAST criteria goals}}
To evaluate our design framework, we determined MAST criteria goals in designing two versions of READIT: a High-MAST version and a Low-MAST version. The High-MAST version was intended to secure the highest possible MAST ratings (i.e., scoring 3 or ``excellent'') by incorporating the necessary features and information transparency levels. For the Low-MAST version, we sought to meet essential system functionality goals with minimal transparency design features, targeting a MAST rating of 1 (``fair'') for each criterion. 

\paragraph*{\textbf{Step 3: Generating READIT 1.0 feature concepts}}
We focused on devising features that allow both experienced and naïve operators to quickly evaluate the underlying data context, algorithmic performance, and algorithmic processes used by READIT to generate its outputs. A research team member (co-author A.P.) initially drafted the feature descriptions, proposing a MAST rating for each one. \emily{Subsequently, our team of computer scientists, industrial engineers, and human systems engineers engaged in a series of discussions and collectively reviewed and reached a consensus on the feature descriptions and their corresponding MAST ratings.}

We summarize a subset of our generated feature concepts for High-MAST READIT 1.0 in  \autoref{tab:MAST_Criteria_2col}, aligned with which MAST criteria they were designed for.
Detailed definitions for each MAST item and corresponding features in High-MAST and Low-MAST versions are provided in \autoref{appendixA}.

    \begin{table*}[!htb]
        \renewcommand{\arraystretch}{1.05} 
        \centering
        \begin{tabularx}{\textwidth}{@{}p{0.4\textwidth}X@{}}
            \toprule
            \textbf{MAST Criterion} & \textbf{High-MAST READIT 1.0 Feature Concept} \\
            \toprule
            \textbf{Sourcing:} Emphasizes the quality and credibility of the system's underlying sources, data, and methodologies. & 
            The \textit{Raw Content} feature showcases raw datasets alongside their summaries (\autoref{fig:rh1}). The \textit{Diagnostics}, \textit{Data Sheet}, and \textit{Date Range} features (\autoref{fig:rh3}) provide information about the training data and metadata. The \textit{Comparison Summary} feature (\autoref{fig:rh2}) distinguishes content from larger organizations versus individual accounts. \\
            \midrule
            \textbf{Uncertainty:} Addresses the impact of uncertainty factors on key analytic judgments. & 
            The \textit{Diagnostics} feature highlights uncertainties in the underlying content, while the \textit{Alternative Summary needed} alert (\autoref{fig:rh4}) notifies users when data integrity issues arise. \\
            \midrule
            \textbf{Distinguishing:} Gauges the system's ability to differentiate underlying intelligence from assumptions and judgments. & 
            The \textit{Diagnostics} feature infers whether content is human- or machine-generated disinformation, helping users assess content quality. \\
            \midrule
            \textbf{Analysis of Alternatives:} Concerns the system's capacity to assess various decision options and provide alternative outcomes when data uncertainty exceeds thresholds. & 
            The \textit{Alternative Summary needed} feature prompts users to generate new summaries using alternative keywords when uncertainty factors are significant (\autoref{fig:rh4}). \\
            \midrule
            \textbf{Customer Relevance:} Addresses the implications of system outputs and its ability to support user needs. & 
            Features such as \textit{Most Influential Content}, \textit{Extractive Summary} (\autoref{fig:rh1}), \textit{Content Shown}, \textit{Diagnostics}, and \textit{Word Cloud} (\autoref{fig:rh3}) provide tailored insights. \\
            \midrule
            \textbf{Logical Argumentation:} Pertains to the system's application of clear and logical reasoning in elucidating the AI's information generation process. & 
            \textit{Extractive Summaries} are ranked by significance scores, and search results are ordered by follower count in the \textit{Summary} and \textit{Raw Content} features. \\
            \midrule
            \textbf{Consistency:} Emphasizes the system's capability to explain the relevance of, or changes in, analytic judgments over time. & 
            The \textit{Comparison Summary} feature provides statistics to compare and contrast summaries across sessions (\autoref{fig:rh2}). \\
            \midrule
            \textbf{Accuracy:} Pertains to the precision of a system's judgments and their clarity. & 
            The \textit{Date Range} and \textit{Diagnostics} features provide metadata, allowing operators to assess content relevance and quality. \\
            \midrule
            \textbf{Visualization:} Describes the integration of effective visual data displays. & 
            The \textit{Visualizations} feature (\autoref{fig:rh2}) includes bar charts, histograms, word clouds, and network graphs for interpretation. \\
            \bottomrule
        \end{tabularx}
        \caption{MAST Criteria and Corresponding Feature Concepts in High-MAST READIT 1.0.}
        \label{tab:MAST_Criteria_2col}
        \renewcommand{\arraystretch}{1.0} 
    \end{table*}

\paragraph*{\textbf{Step 4: Generating READIT 1.0 system mockups}}
After selecting a final set of MAST-informed feature concepts, we created system mockups of High-MAST and Low-MAST versions. The High-MAST READIT 1.0 mockup (Figures \ref{fig:rh1}-\ref{fig:rh4}) comprises the following:

\begin{enumerate}
    \item The \textit{Main Page} (\autoref{fig:rh1}), featuring a keyword search bar alongside summaries and source texts;
    \item  The \textit{Tools} window (\autoref{fig:rh2}), \emily{accessible via a button on the main page, offering advanced search options and content visualizations;} and
    \item The \textit{Options} (\autoref{fig:rh3}) window, presenting data specifics such as date range, dataset size, and content flagged for false information, special characters, or misspellings.
    \item The \textit{Alternative Summary Needed Alert}, which prompts users to consider alternative summaries when search results surpass a predefined threshold (\autoref{fig:rh4}). 
\end{enumerate}

The Low-MAST READIT 1.0 system mockup (\autoref{fig:MAST_rl}) consists of a single window comprising only essential features necessary for task completion: (1) a keyword search bar; and (2) a display area for search results showcasing the most influential content and their extractive summaries.

To guide our mockup design efforts, we devised a fictitious scenario wherein an intelligence analyst must monitor social media posts and report security-related information regarding public sentiment on the construction of a new U.S. consulate building in London.

\afterpage{%
    \begin{sidewaysfigure*}[htbp]
        \centering
        \includegraphics[width=0.9\textheight]{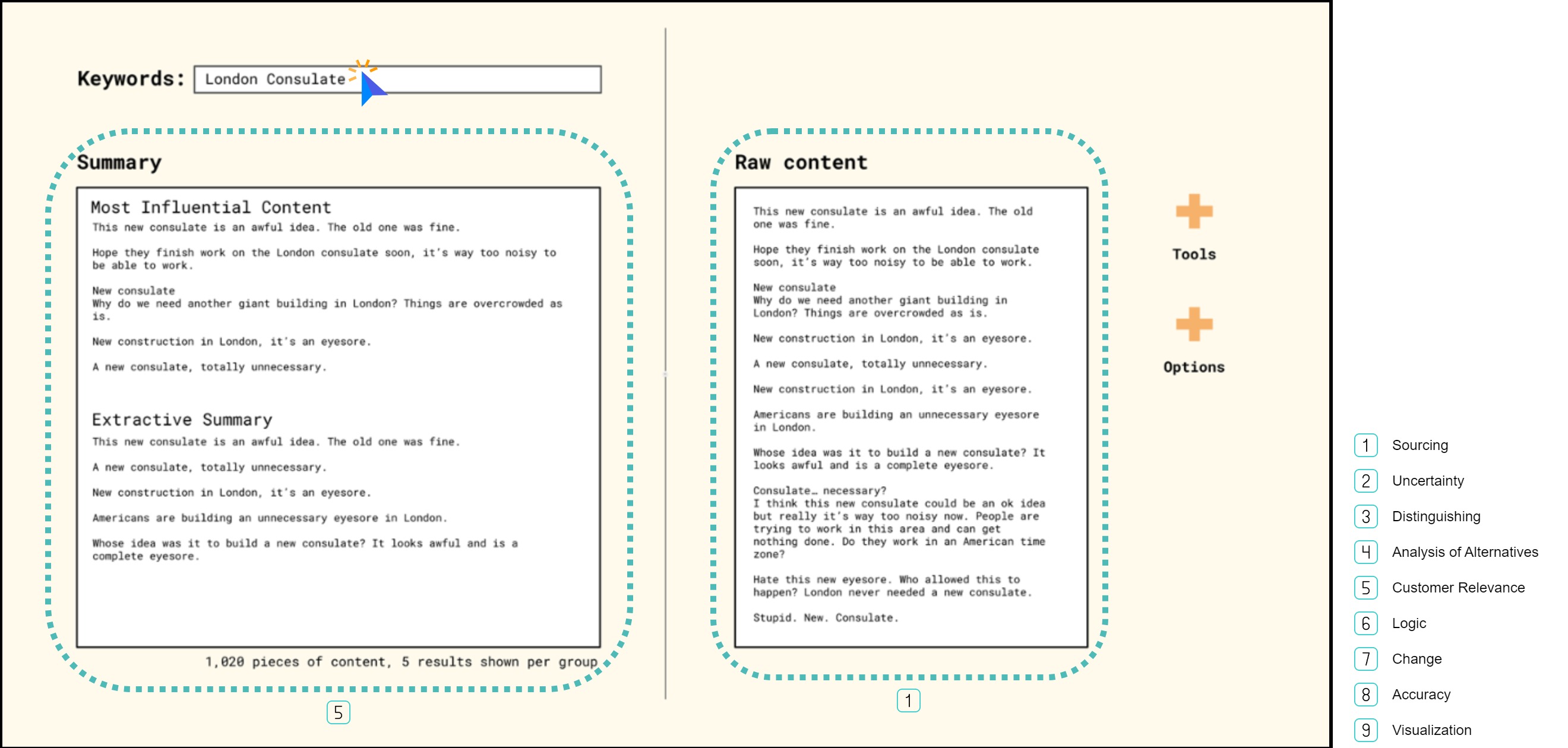}
        \caption{Main page of High-MAST READIT 1.0.}
        \label{fig:rh1}
    \end{sidewaysfigure*}

    \begin{sidewaysfigure*}[htbp]
        \centering
        \includegraphics[width=0.9\textheight]{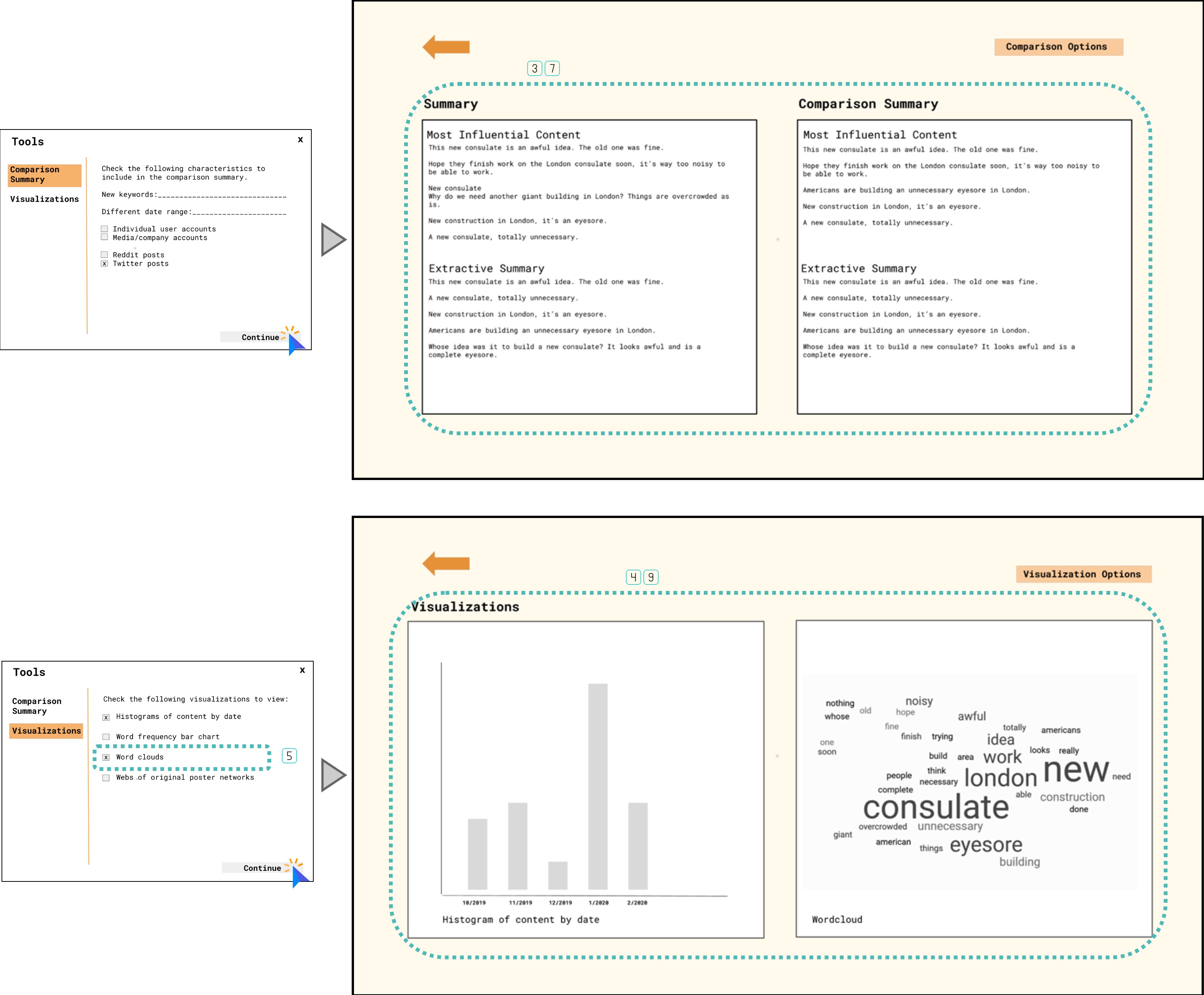}
        \caption{\textit{Tools} window of High-MAST READIT 1.0.}
        \label{fig:rh2}
    \end{sidewaysfigure*}

    \begin{figure*}[htbp]
        \centering
        \includegraphics[width=0.9\textwidth]{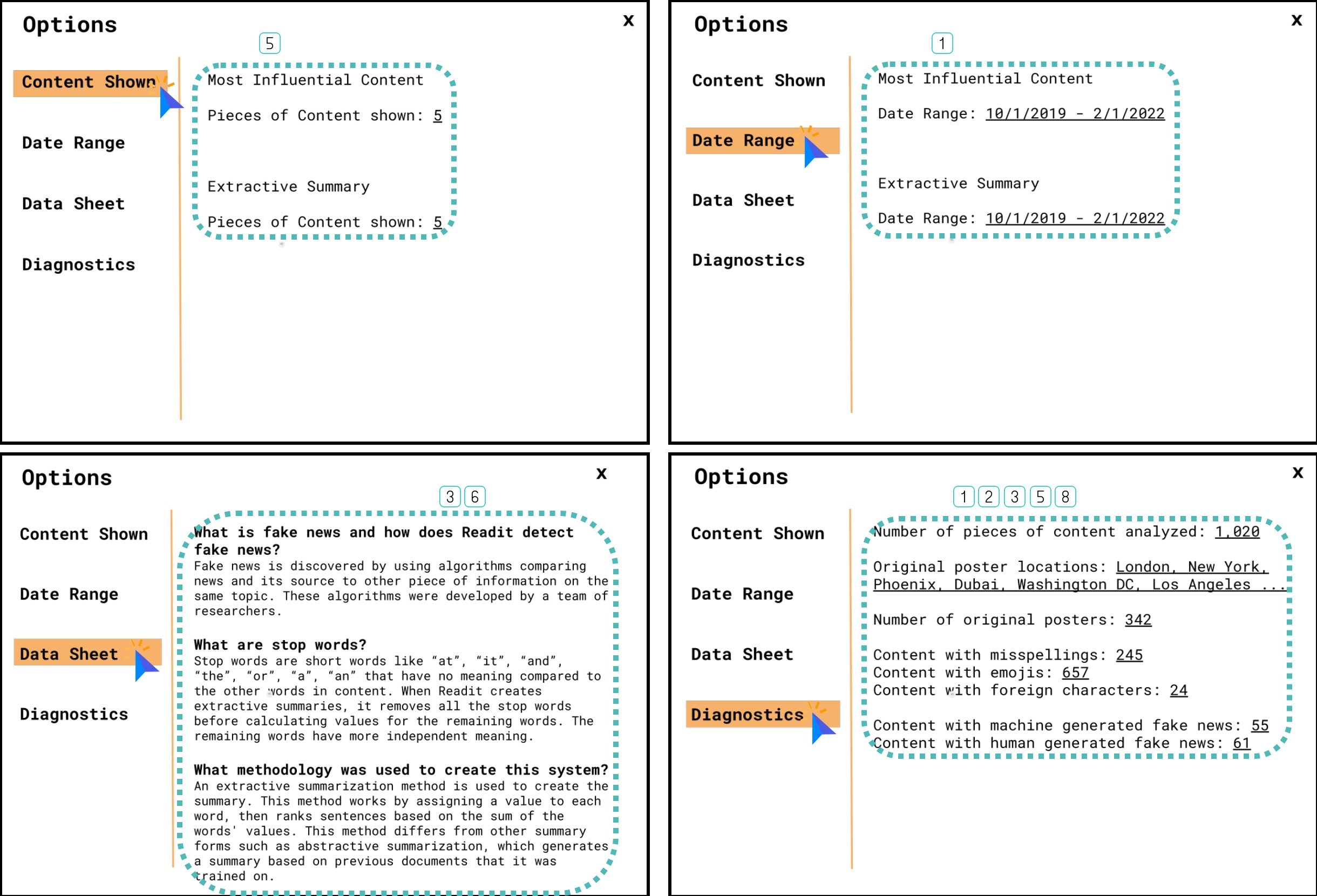}
        \caption{\textit{Options} windows of High-MAST READIT 1.0.}
        \label{fig:rh3}
    \end{figure*}
    
    \begin{figure*}[htbp]
        \centering
        \includegraphics[width=0.9\textwidth]{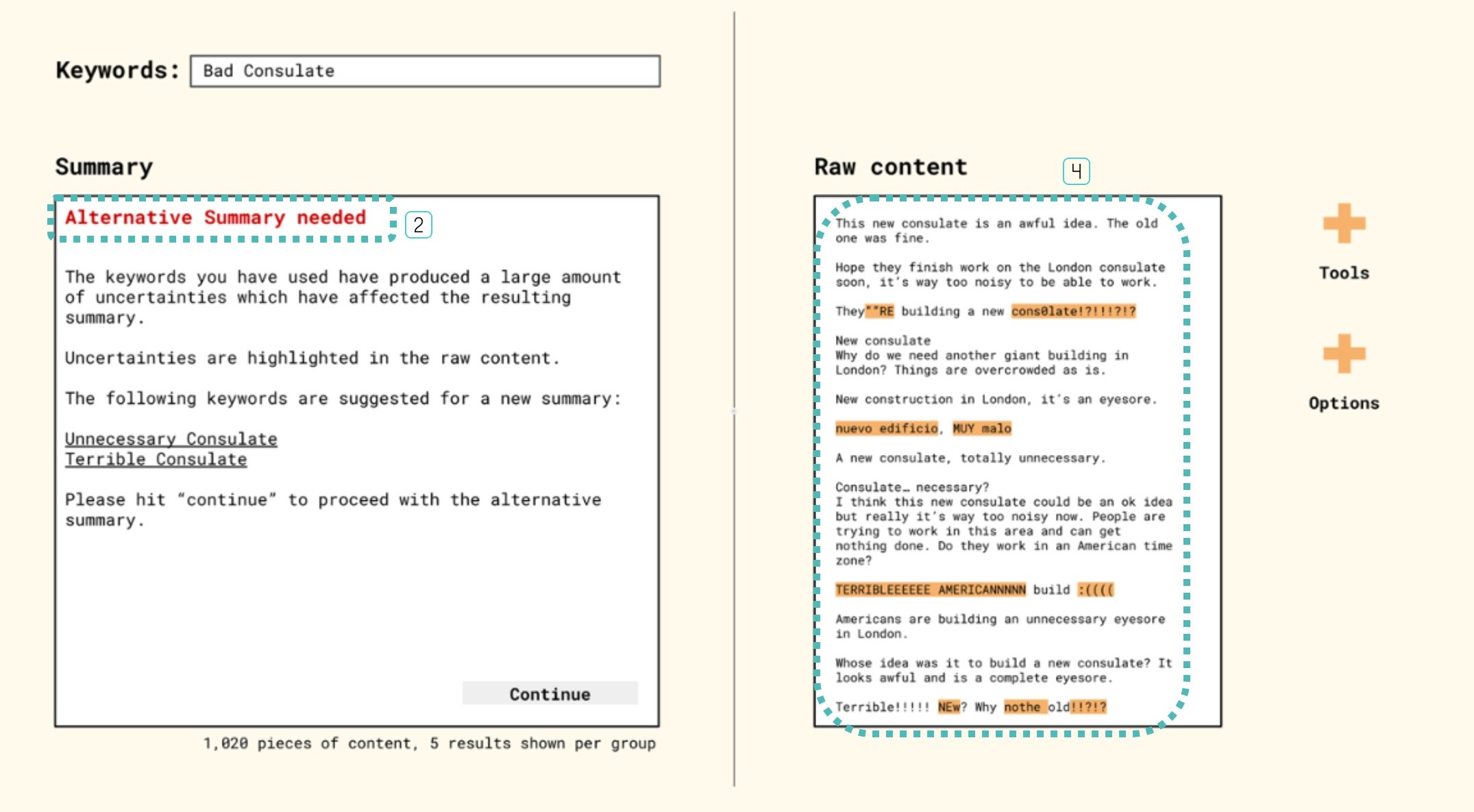}
        \caption{Alternative Summary needed alert in High-MAST READIT 1.0. Highlighted on the ``Raw Content'' screen are text features contributing to high output uncertainty.}
        \label{fig:rh4}
    \end{figure*}
    \FloatBarrier
}

\afterpage{%
    \begin{figure}[htbp!]
        \centering
        \includegraphics[width=0.45\textwidth]{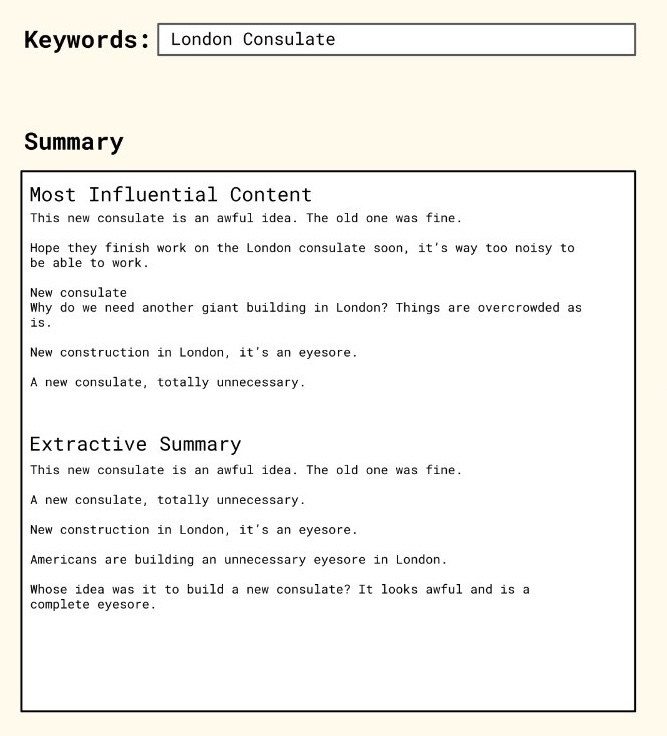}
        \caption{Example search result of Low-MAST READIT 1.0, demonstrating the same essential system functionalities as its High-MAST counterpart.}
        \label{fig:MAST_rl}
    \end{figure}          
}

\paragraph*{\textbf{Step 5: Determining READIT 1.0 MAST ratings}}
As an initial evaluation of the effectiveness of PADTHAI-MM's concept generation steps, we performed Step 5 through an experimental study that aimed to establish and compare end-user ratings for High-MAST and Low-MAST READIT 1.0 system mockups (\autoref{fig:flowchart}).

\afterpage{%
    \begin{figure}[htbp]
    \centering
    \includegraphics[scale=0.5]{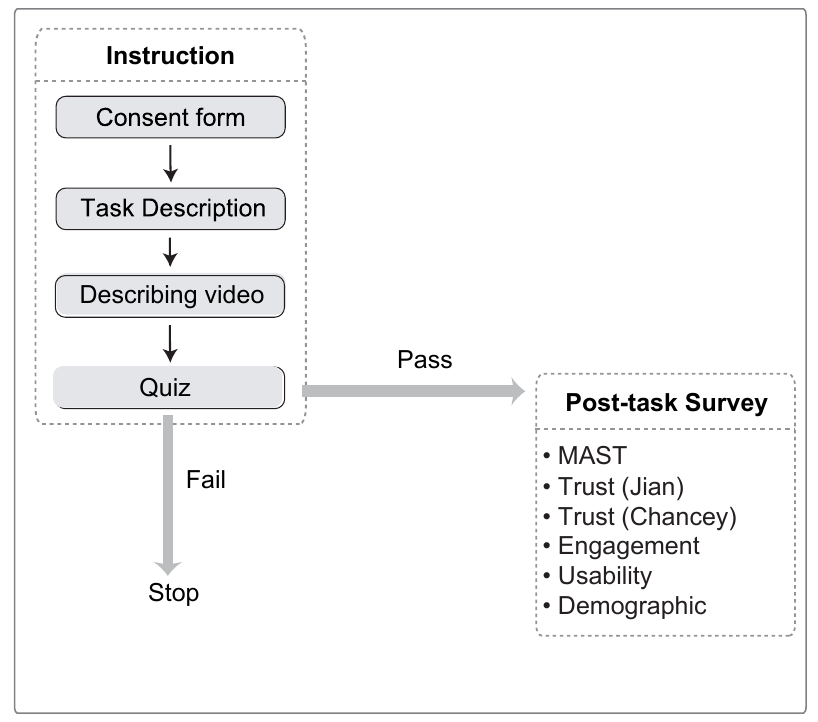}
    \caption{Evaluation Study Procedure for READIT 1.0.}
    \label{fig:flowchart}
    \end{figure}
}


\textbf{\textit{Participants.}} We report data from 43 participants recruited through Prolific (www.prolific.com), of whom 26 were assigned to the Low-MAST condition and 17 to the High-MAST condition.

\textit{\textbf{Dependent Variables: MAST and Trust Surveys.}} We gathered MAST ratings using an adaptation of the \cite{blasch2021multisource} four-point MAST scale, with ``1'' representing ``poor'' and ``4'' representing ``excellent''. After initial pilot testing, our team decided to rephrase each MAST criterion as a question accompanied by its relevant feature description, as shown in \autoref{appendixB}. 
The MAST-total score was calculated by adding these 9 criteria with a min - max range of 9 - 36. 
We also administered two well-known questionnaires used to assess people's trust in automation, with 12 items \citep{jian2000foundations} assessing general trust and 15 items \citep{chancey2017trust} assessing three types of information known to affect trust in automation: performance, process, and purpose. Each item was rated on a seven-point Likert scale, with ``7'' indicating ``extremely agree'' and ``1'' indicating ``extremely disagree''. We calculated the average rating, including the average from each subcategory in \cite{chancey2017trust}, to form the trust perception scores.
    
\textit{\textbf{Control Variables: Engagement and Usability.}} To account for potential confounding factors that could affect the MAST and trust ratings for READIT, we also measured participants' perceptions of their respective READIT 1.0 version's usability using a 10-item questionnaire \citep{brooke2020sus}, as well as their perceived engagement during the evaluation task using a 17-item questionnaire \citep{schaufeli2002measurement}.

\textit{\textbf{Procedure.}} To access the study, participants were directed to a Qualtrics survey link \citep{Qualtrics}. Upon providing their Prolific IDs in the first page, participants were asked to review an IRB-approved informed consent form. Consenting participants were then directed to a task description page that asked them to imagine a fictional scenario in which they were U.S. intelligence analysts, \emily{whose} main task was to monitor social media platforms for ongoing threats to public safety. They were told that to aid them in this task, they would be using an AI-enabled decision support system (i.e., READIT), a technological aid meant to supplement their own skills. Participants were informed that they would not be able to interact with READIT, but that they would be presented with an audio-visual walk-through and description of the tool's interface and features. To check for engagement and comprehension, participants were asked to complete a short pass/fail quiz relating to the content of the task and READIT descriptions. Participants who did not pass were removed from the remainder of the study. Lastly, participants were asked to complete an evaluation of the system using MAST and respond to questionnaires with items measuring trust, task engagement, and usability.

\textit{\textbf{Results.}} Study results showed that the High-MAST group compared to the Low-MAST group was significantly different (descriptively higher) on the MAST ratings (on 8/9 criteria, except \textit{Customer Relevance}) and also significantly different (descriptively higher) on the trust responses. 
In addition, usability and engagement were not significantly different between High-MAST and Low-MAST groups. 


\paragraph*{\textit{\textbf{Step 6: Iterating READIT feature and system concepts}}}
\emily{After analyzing the study data, our team assessed the READIT 1.0 system concepts by considering user evaluations from Step 5 and evaluating the feasibility of implementation in light of our available resources (e.g., datasets, servers, project personnel, timeline, and state-of-the-art models with similar functions).} Although system concepts met the MAST criteria goals, feedback from pilot participants and project stakeholders strongly suggested that READIT 1.0 was not implementable as a functional or credible working prototype for testing with actual intelligence analysts in subsequent steps. We thus redesigned the use case and system as \textbf{READIT 2.0}.

\textbf{\textit{READIT 2.0 Use Case.}} A main constraint with the original use case (i.e., gauging public sentiment from social media) was that it could not be tested with an appropriate dataset to support the analysis and data visualization features we envisioned; generating our own datasets would have gone beyond the scope of our project. We adapted this original use case for compatibility with an existing dataset from the VAST Challenge 2011 MC3 \citep{IEEEVAST2011MC3}: Investigation into Terrorist Activity. In this scenario, participants play the role of intelligence analysts searching for information related to potential terrorist activity among thousands of fictitious news reports. Consultation with project stakeholders confirmed that this revised task sufficiently aligned with real-world situations encountered by analysts, who are trained to detect misinformation but require machine aid to do so at scale.

\begin{sidewaysfigure*}[htbp!]
    \centering
    \includegraphics[width=0.9\textheight]{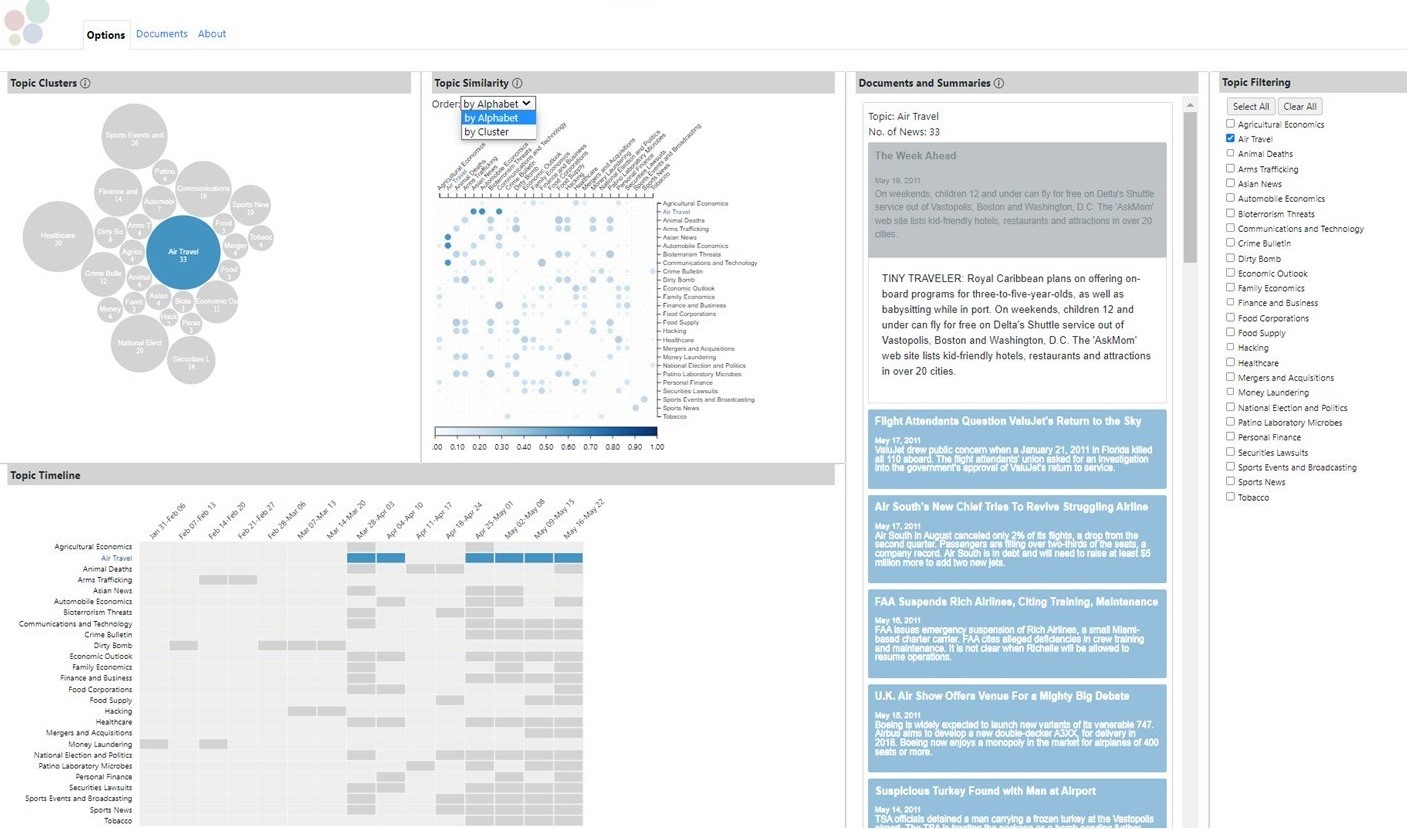}
    \caption{Main page of High-MAST READIT 2.0. Compared to the Low-MAST version, High-MAST includes more interactive features such as \textit{Topic Clusters}, information buttons, and a dropdown option in \textit{Topic Similarity}. It also provides additional information, including the \textit{Documents} and \textit{About} tabs, to better demonstrate the MAST criteria.}
    \label{fig:newrh}
\end{sidewaysfigure*}

\begin{sidewaysfigure*}[htbp!]
    \centering
    \includegraphics[width=0.9\textheight]{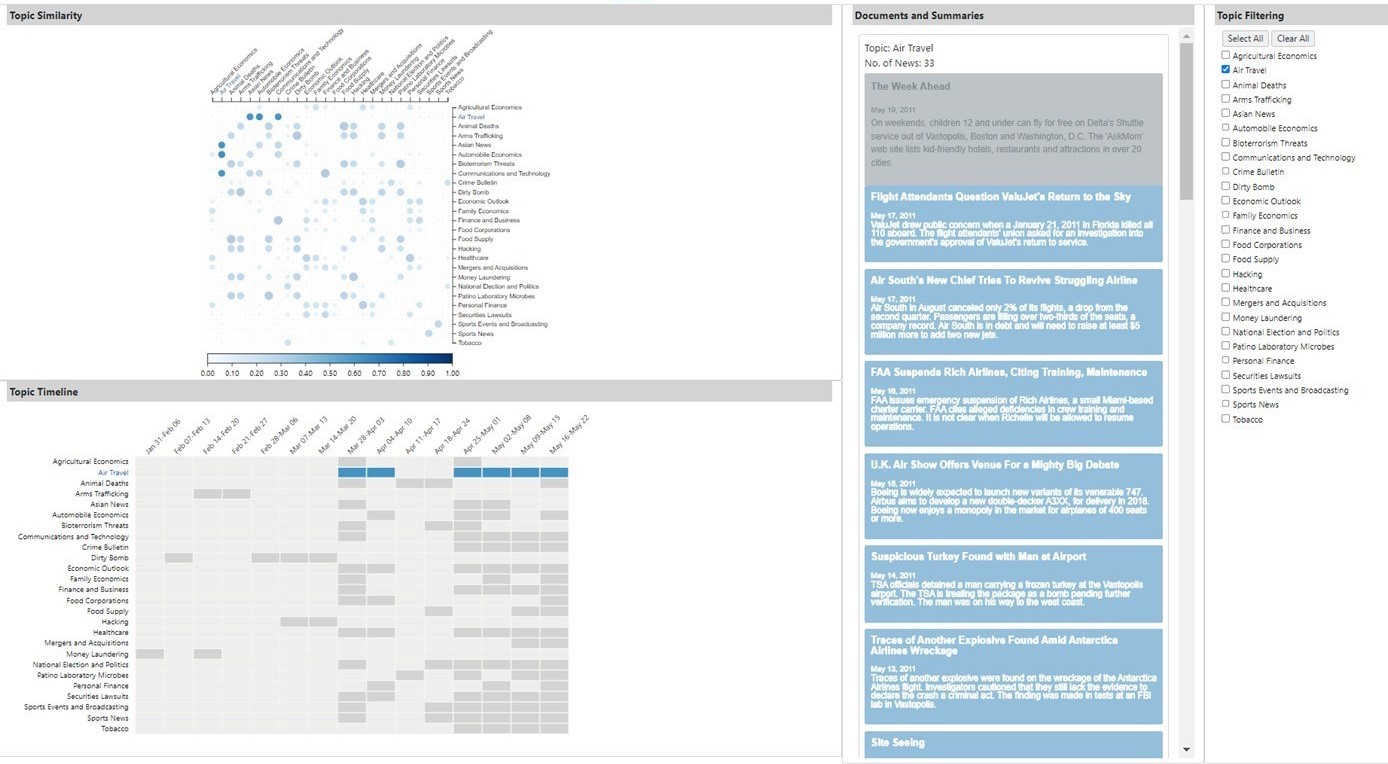}
    \caption{Main page of Low-MAST READIT 2.0. This version lacks some of the interactive features and additional information found in the High-MAST version, demonstrating a minimal configuration of MAST criteria.}
    \label{fig:newrl}
\end{sidewaysfigure*}

\textbf{\textit{READIT 2.0 Feature and System Concepts.}} We again developed Low-MAST and High-MAST versions for READIT 2.0. Instead of the extractive summaries that select and rearrange existing content from original text sources, we designed READIT 2.0 to present \textit{abstractive} summaries, which present novel AI-generated text descriptions of news data. For the High-MAST version (\autoref{fig:newrh}), we designed three data visualization features: \textit{Topic Clusters}, a \textit{Topic Similarity Matrix}, and a \textit{Topic Timeline}. A feature was also designed to provide users with \textit{raw content} when operators click on the summary box, as well as a list of all topics along with checkboxes to narrow down the range of summary results. The Low-MAST version (\autoref{fig:newrl}) provides neither \textit{Topic Clusters} nor additional information buttons for each section, but maintains basic functionality for operators to complete the given task within a reasonable time frame.


After designing the READIT 2.0 mockups, our team assessed the two designs to ensure that the MAST rating goals would be met (``excellent'' for High-MAST, ``fair'' for Low-MAST). \autoref{tab:readit_mast} presents the results of this process, in which each rating represents the outcome of a group discussion and is presented as a binary rating (yes/no) to support efficient decision-making and minimize ambiguity. To fulfill MAST criteria goals, additional features were added to the High-MAST version, including information about the platform and technologies used, as well as a visualization of the topics from the raw documents. To further streamline the Low-MAST version, certain components were simplified, resulting in some functions being omitted. For example, the Low-MAST \textit{Topic Similarity matrix} lacked a selection feature, the \textit{Topic Timeline} lacked a timeline selection option, and the \textit{Documents \& Summaries} did not visualize the documents (presenting only a summary). \autoref{appendixB} presents the detailed features in READIT 2.0 matching each MAST criterion. 

    \begin{table*}[htbp!]
    \tiny
    \centering
        \begin{tabular}{p{0.15\linewidth} c c c c c c c c}
          \toprule 
           & \textbf{\makecell{Data\\Preprocess}} & \textbf{\makecell{Documents\\Tab}} & \textbf{\makecell{About\\Tab}} & \textbf{\makecell{Documents\\ \& Summaries}} & \textbf{\makecell{Topic\\Clusters}} & \textbf{\makecell{Topic\\Similarity\\Matrix}} & \textbf{\makecell{Topic\\Filtering}} & \textbf{\makecell{Topic\\Timeline}} \\ 
          \midrule 
          \textbf{Sourcing} & \checkmark \bluecheckmark & \checkmark & \checkmark & \checkmark & \checkmark & & & \checkmark \\
          \midrule 
          \textbf{Uncertainty} & & & \checkmark & & & \checkmark  &  \\
          \midrule 
          \textbf{Distinguishing} & & \checkmark & \checkmark & \checkmark & \checkmark & \checkmark & \checkmark  & \checkmark \\    
          \midrule 
          \textbf{Analysis of Alternatives} & & & \checkmark & & \checkmark & \checkmark & \checkmark & \\
          
          \midrule 
          \textbf{Customer Relevance} & & & & \checkmark & \checkmark & \checkmark \bluecheckmark & \checkmark \bluecheckmark & \checkmark \bluecheckmark\\                            
          \midrule 
          \textbf{Logic} & & & \checkmark & \checkmark \bluecheckmark & \checkmark & \checkmark \bluecheckmark & \checkmark & \checkmark\\
          
          \midrule 
          \textbf{Change} & & \checkmark & \checkmark & & & \checkmark &  \\
          \midrule 
          \textbf{Accuracy} & & & \checkmark & & \checkmark & & & \\
          \midrule 
          \textbf{Visualization} & \checkmark \bluecheckmark & & & \checkmark & \checkmark & \checkmark \bluecheckmark & \checkmark & \checkmark \\        
        \bottomrule 
        \end{tabular}
        \vspace{20pt} \caption{MAST criteria in READIT 2.0. Red checkmarks and blue checkmarks denote High-MAST and Low-MAST versions respectively. Each row represents a specific MAST criterion and the columns show details for the front-end interface, which include various functions presented by the AI system, such as descriptive and calculated details, alerts, pop-up boxes, and buttons.}
      \label{tab:readit_mast}
    \end{table*}

\paragraph*{\textbf{Step 7: Generating READIT 2.0 working prototypes}}
READIT 2.0 was prototyped as a JavaScript-based client to facilitate interactivity between participants and the AI-DSS. The server was developed using Python 3 and Flask library and hosted on the Google Cloud Platform. To enhance operators' ability to explore the dataset, READIT 2.0 included various data visualizations created using the React JavaScript library.

We designed the abstractive summary generation feature of READIT 2.0 using PEGASUS \citep{zhang2020pegasus}, a state-of-the-art deep learning-based text summarization algorithm. The model uses the transformer encoder-decoder structure. Input text is encoded as a numerical context vector, which is then decoded to generate the abstractive summaries. The model is pre-trained to choose and mask important sentences and then recover them, a process called Gap Sentence Generation. We used the cnn\_dailymail dataset \citep{seeGetPointSummarization2017} for pre-training PEGASUS. Both High-MAST and Low-MAST working prototypes of READIT 2.0 included a functional front-end interface that allowed access to various natural language processing and visualization features executed in the back-end. The code is available in \cite{kimNayoungkim94PADTHAIMM2023}.

\paragraph*{\textbf{Step 8: Evaluating READIT 2.0 Working Prototypes}}
As in Step 5, we designed a remote experimental protocol to evaluate user perceptions of READIT 2.0 working prototypes. In this step, we also sought to assess the viability of the PADTHAI-MM framework through comparing Low- and High-MAST versions, including how participants' decision-making metrics. We summarize the results of this experiment, first reported in detail by \cite{salehiTrustworthyAIEnabledDecision2024}.

We recruited 23 intelligence analysts (IAs) from the U.S. Department of Homeland Security (DHS), who were considered subject domain experts and were randomly assigned to either a High-MAST or Low-MAST version and tasked with using READIT in a hypothetical scenario aimed at identifying potential \nk{patterns} of a terrorist attack. Afterward, participants were asked to assign MAST criteria ratings and trust questionnaires developed by \cite{jian2000foundations} and \cite{chancey2017trust}.

Our findings showed the High-MAST group exhibited descriptively higher ratings across 8 of the 9 MAST criteria, with the exception of \textit{Customer Relevance}, and displayed significantly higher trust ratings in the ``purpose'' dimension \citep{chancey2017trust} compared to the Low-MAST group.
Concerning usability and engagement, no significant differences were observed between the groups. Notably, the High-MAST group required an average of 10 additional minutes to complete the task relative to the Low-MAST group. 

\paragraph*{\textbf{Step 9: Determining READIT 2.0 further iteration needs}}
Following the examination of participant feedback on READIT 2.0, our design team concluded that the system successfully fulfilled our specified functionality and MAST criteria objectives. Given that the primary aim of this project was to facilitate research, we decided against further design iterations.

%% file: sections/4.2_Validation.tex
\subsection{PADTHAI-MM Theoretical Validation}
In addition to demonstrating the practical utility of the PADTHAI-MM framework, we investigated whether system features designed using the MAST criteria aligned with participants' trust perceptions across three trust constructs---namely, \textit{Process}, \textit{Purpose}, and \textit{Performance} \citep{lee2004trust}. MAST ratings on system features were derived with two key objectives: (1) to validate whether MAST-informed design principles translated into participants' perceived trust; and (2) to identify which system features strongly supported each trust construct, thus providing actionable insights for designing trustworthy AI systems.

\subsubsection{Method}

Because we specifically used the nine MAST criteria to design the two READIT versions, the first step was for members of our team (authors NK, YB, SB, PS, EC, and MM) to theoretically determine, through voting and group consensus, if a MAST sub-item was related to an information-based construct of trust (process, purpose, performance). As an example, the MAST sub-item \emph{``System analyzes in detail how the current derived results relate to other information of interest to the user"} was unanimously assessed as being related system to \emph{Purpose}. For sub-items where a unanimous consensus was not reached, we adopted an inclusive approach, assigning the sub-item to all constructs that received at least one vote. For instance, the sub-item ``The system highlights data in an effective manner that clarifies, complements, or augments derived results and enhances user understanding" received 5 votes for \emph{Purpose} and 1 for \emph{Process}, so the sub-item was included in both constructs. This process resulted in MAST ratings for related trust constructs in the form of matrices $\boldsymbol{\mathrm{P_i}}_{(m\times n_i)}, i = 1, 2, 3$ (see Appendix \ref{appendixE} for details). 

Each matrix $\boldsymbol{\mathrm{P_i}}$ was multiplied to the MAST-system matrix $\boldsymbol{\mathrm{F}_{(n_i\times p)}}$, which indicates the presence (1) or absence (0) of specific MAST sub-items addressed by each system feature $j = 1, 2, \ldots, p$. \autoref{tab:readit_mast} shows the mapping of general system features with the MAST criteria. The resulting matrix, $\boldsymbol{\mathrm{T_i = P_i \times F}}$, provides construct-specific MAST ratings of the system features. Each matrix $\boldsymbol{\mathrm{T_i}}$ corresponds to one of the three trust constructs (i.e., Process, Purpose, or Performance) and evaluates how well system features align with the MAST criteria for the given construct.

Principal Component Analysis (PCA) was applied to each $\boldsymbol{\mathrm{T_i}}$ to identify the system features most strongly associated with each trust construct (process, purpose, or performance) \citep{abdiPrincipalComponentAnalysis2010}. For the PCA, we established two criteria: (1) the initial $k$ principal components should account for at least 80\% of the cumulative variance in the data; and (2) factor loadings should have absolute values of at least $0.3$.

As a final step, we conducted a series of linear regressions using participants' average trust ratings for the three constructs (process, purpose, and performance) as dependent variables. These ratings are represented as matrices \( \boldsymbol{\mathrm{TR_i}} \), where \( i = 1, 2, 3 \) corresponds to each trust construct. The independent variables were the principal component (PC) scores derived from each \( \boldsymbol{\mathrm{T_i}} \). This modeling was performed in R, employing the \textit{dplyr}~\citep{wickhamr} and \textit{psych}~\citep{revelle2018procedures} packages. The mathematical representation of the regression model is as follows:

\begin{equation}
    \boldsymbol{\mathrm{TR_i}} =
    \boldsymbol{\beta_i}
    \begin{bmatrix}
    PC_1 \\
    PC_2 \\
    \vdots \\
    PC_k
    \end{bmatrix}
    + 
    \boldsymbol{\varepsilon_i}
\end{equation}

\noindent where:
\begin{itemize}
    \item \( \boldsymbol{\mathrm{TR_i}} \in \mathbb{R}^{m \times 1} \): The \( m \times 1 \) matrix of participants' trust ratings for construct \( i \), where \( i \) corresponds to process, purpose, or performance.
    \item \( \boldsymbol{\beta_i} \in \mathbb{R}^{k \times 1} \): The vector of regression coefficients for the \( k \) principal components associated with \( \boldsymbol{\mathrm{T_i}} \).
    \item \( PC_k \in \mathbb{R}^{m \times 1} \): The \( m \times 1 \) matrix of scores for the \( k^{\text{th}} \) principal component derived from \( \boldsymbol{\mathrm{T_i}} \), where \( k \) indexes the retained components accounting for at least 80\% variance.
    \item \( \boldsymbol{\varepsilon_i} \in \mathbb{R}^{m \times 1} \): The residual error matrix for trust construct \( i \).
\end{itemize}

\subsubsection{Results}

\begin{table*}[!htbp]
    \centering
    \begin{tabularx}{\textwidth}{>{\raggedright\arraybackslash}X >{\centering\arraybackslash}p{0.15\linewidth} >{\centering\arraybackslash}p{0.15\linewidth} >{\centering\arraybackslash}p{0.2\linewidth}}
      \toprule
      \multicolumn{1}{c}{\textbf{READIT 2.0 System Features}} & \textbf{\textit{Process}} & \textbf{\textit{Purpose}} & \textbf{\textit{Performance}} \\ 
      \toprule
      Data Preprocessing &  &   & \\
      \midrule
      Documents Tab &  &   & \\
      \midrule
      About Tab & 0.477 & 0.68 & \\
      \midrule
      Topic Clusters & 0.287 & 0.328 & 0.479 \\
      \midrule
      Topic Similarity Matrix & 0.314 & 0.464 & 0.51 \\
      \midrule
      Topic Filtering &  & 0.412 & \\
      \midrule
      Topic Timeline &  & 0.464 & 0.388 \\
      \midrule
      Documents and Summaries & 0.246 & 0.328 & \\
      \bottomrule
    \end{tabularx}
    \vspace{20pt}
    \caption{Correlation between READIT 2.0 features and process, purpose, and performance information dimensions. The values in the table represent the nested coefficients obtained from PCA loadings multiplied by regression coefficients $\beta$. (Detailed information about loadings and coefficients are found in \autoref{appendixD}.) Blank spaces indicate that certain features do not meet the 0.3 threshold of PCA loadings, suggesting they are not as significant as other features for Process, Purpose, and Performance.}
\label{tab:pca}
\end{table*}

\emily{\autoref{tab:pca} shows the correlation between READIT 2.0's features and each trust-related information dimension, illustrating the expected shift in \cite{chancey2017trust} subscale ratings corresponding to a one-standard-deviation increase in the participants' ratings of MAST-informed features.} For instance, a standard deviation increase in the relevant MAST ratings for the About Tab enhances the \emph{Purpose} rating by an average of 0.68 units, assuming all other predictors remain constant. 

PCA findings reveal that the integration of MAST-informed features within READIT 2.0 demonstrates a moderate to strong correlation with each of the three trust-related information dimensions, indicating a significant influence of PADTHAI-MM's MAST-inspired approach to influencing system trustworthiness. However, the impact of dominant features on \textit{Process}, \textit{Purpose}, and \textit{Performance} is varied, with multiple features of READIT 2.0 significantly contributing to all three. For instance, \textit{Process} was largely associated with the About Tab, Topic Clusters, Topic Similarity Matrix, and Documents and Summaries. In contrast, \textit{Purpose} was broadly addressed by all features with the exception of Data Preprocessing and Documents Tab. The About Tab emerges as particularly crucial in this context, possibly because the availability of detailed insights can allow users to perceive having a deeper understanding of READIT 2.0’s functionalities.
\textit{Performance} was notably specifically enhanced by the synergistic features of \textit{Topic Clusters}, \textit{Topic Similarity Matrix}, and \textit{Topic Timeline} when a specific topic is selected. \emily{These features together provide a comprehensive and structured understanding of the data through visualization techniques, thereby aiding operators in navigating and using the data more effectively.}

This analysis suggests that PADTHAI-MM enabled the development and selection of READIT 2.0 features that addressed the critical dimensions of process, purpose, and performance for this specific use case. We consider this a small, but crucial, evidence for the theoretical validity of the MAST-inspired PADTHAI-MM framework within established scholarship on users' trust in automation.

%% file: sections/5_Discussion.tex
\section{Discussion}


Our primary goal for this work was to demonstrate how tradecraft-derived standards and tools can be adapted into accessible design guidelines for creating AI-DSSs that address context-specific information needs to assess system trustworthiness. PADTHAI-MM is a step-by-step framework for designing AI systems according to established principles within the intelligence community, based on Blasch et al.'s (\citeyear{blasch2021multisource}) MAST AI-DSS evaluation tool. Applying PADTHAI-MM for our design case study resulted in two prototypes of READIT, a text summarization tool: a High-MAST version, enriched with a comprehensive feature set aimed at augmenting operator task performance, and a Low-MAST version, streamlined to include only the essential features needed for task completion. Statistically significant differences between stakeholders' evaluations of High-MAST and Low-MAST versions, both using MAST ratings and \cite{chancey2017trust} responses, support the effectiveness of PADTHAI-MM as a framework for system trustworthiness as a design objective. This is further supported by the significant relationships between MAST-informed features and participant perceptions of a system \textit{process}, \textit{purpose}, and \textit{performance}. Overall, these results not only validate the practical application of MAST within a design context but also suggests that a MAST-informed design process can support the design and development of more trustable AI-DSSs.

PADTHAI-MM shares similar principles\emily{ to those} behind user-centered and work-centered design approaches for software systems (e.g., \citealp{Beyer_Holtzblatt_1999, Roth_Bisantz_Wang_Kim_Hettinger_2021}). However, many user-centered approaches tend to broader focus on operator goals and specifications that could improve technology adoption, and work-centered approaches tend to focus on reducing cognitive load\emily{—}both crucial for broader system design considerations. PADTHAI-MM focuses more simply on how to incorporate organizational values and state-of-the-art knowledge of technology capability with criteria for information quality that are widely adopted by the Intelligence Community. For example, throughout the Iterative Evaluation steps (i.e., Steps 6 and 9), system design iterations attuned to the MAST criteria are generated after stakeholder inputs to address intelligence community trustworthiness needs and expectations. PADTHAI-MM outlines a principles-based iterative design process that can aid system developer teams simultaneously meet system functionality goals and domain-specific trustworthiness needs---even those whose primary expertise areas are in AI and data system concepts (e.g., principles, tools, and processes linked to the datasets that will be used by the AI-DSS), and not in product design or human factors.

For this study, we combined an intelligence analytics task scenario and an NLP text summarization use case to maximize ecological validity, considering MAST's foundations in intelligence analytics standards (i.e., ICD 203; \citealp{odniIntelligenceCommunityDirective2015}) and tool development within the NLP domain \citep{blasch2021multisource}. However, MAST has been shown as an effective AI trustworthiness evaluation tool for applications in air traffic control \citep{insaurraldeTrustEvaluationOntological2021}, satellite imaging anomaly detection \citep{blasch2021b}, cyberattack awareness and response \citep{blaschHumanmachineCooperativeAI2023}, and facial verification technologies \citep{salehiTrustworthyAIEnabledDecision2024}, suggesting the practical relevance of the MAST criteria (and by extension, PADTHAI-MM) beyond intelligence analytics use cases. Our prior research \citep{salehiTrustworthyAIEnabledDecision2024} also suggests that MAST has convergent validity with theoretically-derived measures of perceived system trustworthiness that are broadly applied (e.g., \citealp{chancey2017trust, jian2000foundations}). Further work is needed to evaluate the generalizability of PADTHAI-MM in designing for trustworthiness in other AI application domains, though the aforementioned developments suggest this is a promising direction.



We acknowledge some limitations in this study. First, our Case Study evaluation findings highlight the perceptual differences between the High-MAST and Low-MAST versions from an operator's perspective. However, higher MAST criteria and trust scores do not necessarily equate to improved task performance outcomes~\citep{salehiTrustworthyAIEnabledDecision2024}. This is consistent with the mixed conclusions in the literature on whether users' perceived trustworthiness ratings are directly and positively correlated to operational efficiency and effectiveness \citep{hancockHowWhyHumans2023}. Second, the fast-paced evolution of AI technologies and their various application areas may require updates or extensions to the MAST methodology to maintain its long-term relevance and efficacy in the system design and evaluation space. For example, could there be a time when it is more widely recognized that some criteria will be difficult to achieve an ``excellent'' rating on, depending on the criticality (e.g., risk, threat) of the task? What then should technology evaluators do with that information? Third, results of our PCA regression analysis suggests synergies and potential redundancies between each individual criterion; however, refining the present PADTHAI-MM framework based on these results is beyond the scope of this study. Future efforts may benefit from considering multicollinearity through PCA or other analytic techniques like Correspondence Analysis (CA) that could be used to consolidate the MAST criteria to enhance the framework's efficiency without compromising its comprehensiveness. Lastly, navigating the trade-offs between competing MAST criteria, such as the tension between model Accuracy and Uncertainty, remains a challenge. Future endeavors particularly on fielded systems could aim to reconcile these conflicts by integrating other impactful design approaches (e.g., ~\citealp{Roth_Bisantz_Wang_Kim_Hettinger_2021}) with PADTHAI-MM. 

Looking forward, addressing the practical challenges and potential biases in evaluating AI systems is essential for preserving the integrity and effectiveness of MAST as part of a design methodology. We posit that documenting the variability inherent in context-specific ratings of the MAST criteria might highlight areas that require enhanced scrutiny, particularly in defining feature concepts more precisely and operationalizing them in alignment with the rating objectives. For instance, selection bias could occur, which would reduce the benefits of having a multidisciplinary design team; this is why we suggest the use of formal decision-making tools and strategies (e.g., \citealp{hauser66house, xiaoComparisonConceptSelection2007}) to mitigate the risks of groupthink. Another challenge lies in the practical evaluation of the criteria, especially in assessing data quality and credibility in real-world cases. Although MAST covers this concern in its \textit{Sourcing} criteria, in READIT 2.0, we modified competition data that had verifiable quality and credibility given the well-scoped task and a known ground truth. However, for more complex AI systems, advanced statistical methods may be required to assess data quality given large corpus data environments where ground truth is virtually unknown \citep{ba2024data}. 
Another approach may involve the integration of adaptive AI architectures and algorithms (e.g., \citealp{madrasPredictResponsiblyImproving2018}) into MAST-inspired system designs to take operator needs and decision biases into consideration when displaying trust-enhancing features or system outputs. For example, algorithmic approaches to detect paralinguistic features like sarcasm (e.g., \citealp{xiaoNeuroInspiredInformationTheoreticHierarchical2024}) may be necessary due to subpar human capabilities when handling data at scale. As MAST continues to be refined, we encourage future researchers, designers, or developers to consider how different AI models and their application contexts might differ from the ones we discuss here, and to familiarize \emily{themselves }with other design approaches to more comprehensively advance the development of trustable and trustworthy AI systems. 

%% file: sections/6_Conclusion.tex
\section{Conclusion}
Trustworthy AI systems support users in appraising AI outputs through context-appropriate transparency and interactivity. Yet, the demands of contextualizing users' trust-related information needs and bridging them with appropriate design solutions remain burdensome to system designer teams—particularly when designing for high-stakes decision domains. In this study, we illustrated how tradecraft-derived standards, tools, and principles can be resources for aligning user-centered trust design considerations with operational requirements. Specifically, we introduced PADTHAI-MM: a step-by-step design methodology that integrates scholarship on human-automation trust, product design and development, and intelligence tradecraft trustworthiness standards. As demonstrated in our READIT design case study, PADTHAI-MM is a functional AI design and evaluation framework for intelligence analytics tasks, with potential applicability in domains with similar established values regarding system trustworthiness. Our approach in creating PADTHAI-MM offers a foundation for future trustworthy and human-centered AI design frameworks that integrate theory and practice with principles outlined in tradecraft resources like the MAST methodology.



%% file: sections/7_Acknowledgments.tex
\section*{Acknowledgments}
This material is based on work supported by the U. S. Department of Homeland Security under Grant Award Number 17STQAC00001-05-00. The views and conclusions contained in this document are those of the authors and should not be interpreted as representing the official policies, either expressed or implied, of the Department of Homeland Security, Department of Defense, or the university.\erin{ Copy edits by Emily Summers improved the final version of this manuscript.}

%% file: sections/appendixA.tex
\newpage 
\appendix
\section{MAST Target-Feature Mapping for READIT 1.0}
\label{appendixA}

\scriptsize

\begin{longtable}{p{0.13\linewidth}|p{0.42\linewidth}|p{0.35\linewidth}}
    \toprule
    \textbf{MAST item} & \textbf{High-MAST} & \textbf{Low-MAST} \\
    \midrule
    \textbf{Sourcing}
        & - The system provides a description of the data that was used to derive the result including agencies, serial numbers, number of reports from an agency, intelligence type, source quality.
        
        - System allows you to click on text in summary and it will bring up underlying report and relevant passages in it.
        
        - Datasheet describes the quality of data used, which data repositories were searched, and data used to train and validate the system/methodology used by algorithm and describes potential biases if applied to data not trained for. 
        
        - User/data owner filters reporting so system does not consider duplicates, derivatives, or summaries.
        
        - Systems has detailed knowledge of agencies, organizations, sources, collection methods, source evaluations and can provide summary of factors affecting quality and credibility of sources used to obtain result.
        
        & - The system will not provide a description of the type of content or number of posts used in the summary or any descriptive information regarding the content’s sourcing.
        
        - The system does not have the ability to create reports about the data.\\
      
        \midrule
      
        \textbf{Uncertainty} 
        & - The system provides a detailed description of the accuracy of its summary with an accuracy score and explanation on how the score was derived.
      
        - The system provides detailed diagnostic information about uncertainties in derived results and why they came about, including problems with structure of raw reporting or inconsistent/contrary information in the reporting.
      
        - System can tell what type of data would be needed to increase confidence in results, i.e. better quality reporting in a certain area would increase accuracy of results. 
      
        & - The system will not provide any information about uncertainties in the underlying content, nor will it provide any information about how those uncertainties may have affected the accuracy of the resulting summary. \\

        \midrule
        \textbf{Distinguishing}
        & - The NLP system identifies structural or semantic indicators in the reporting that, if detected, could validate or refute assumptions used to develop the model.
      
        - The developer includes a datasheet or factsheet that details aspects of the training data that were critical to the development of the model, how the system could react to use in unfamiliar environments, and suggestions for how to reduce risks of bias and error.
        
        - The developer of the system includes detailed information that explains how press and government reporting are different, how the model might react, and remedies to be taken to account for the differences in the two types of reporting.

        - The datasheet has information on assumptions that framed how data was chosen, curation rationale, annotations, and collection process.
        
        - There is a model card showing type of model to understand the type of assumptions, domain-relevant expert input/judgment, and current model version.
        
        - There is a factsheet on explainability, such as statement of purpose, if there is direct/indirect use of data and algorithm, and recommended use, users, and restrictions.
        
        - System can explicitly state assumptions it used to fill in gaps when deriving summaries from limited or noisy reporting.

        & - The system does not provide any indicators to help distinguish between derived results (summary) and underlying data (individual pieces of content or posts). For example, there is no way to determine which parts of the summary are from different sources. The whole summary could be from a single or source, or many sources. This is not distinguishable.
      
        - There is no additional information describing how the system was trained to analyze results. Also, there is no information on the specific sources or type of content that was used in the system’s training algorithm. \\
      
        \midrule 
        \textbf{\makecell{Analysis of \\Alternatives}}
        & - The system presents alternative possibilities for results when uncertainties, lack of data, etc. warrant alternatives, indicates which alternative is best, and likelihood that alternatives are correct.
      
        - The system explains evidence and reasoning that underpin the alternative by providing a list of source documents and highlighting most relevant text in alternative summary.
      
        - System indicates which data or keywords which, if detected, affect the likelihood of any identified alternatives.

        & - The system does not suggest any alternative summaries, even if uncertainties in the content warrant an alternative summary. \\
      
        \midrule 
        \textbf{\makecell{Customer \\Relevance}}
        & - Identifies trends or prospects for follow-up work by analyst/user.
      
        - System provides useful information and results that are tailored to needs of analyst/user, can provide supplementary information (sourcing, visuals, etc.) if needed.
      
        - System adds value by synthesizing large volumes of information across disciplines to draw conclusions about trends/prospects for additional analysis, connections between data, classification of data.
      
        - System can also show trends in reporting (use timelines or visuals), extract entities or show connections between them, generate written reports based on analysis of data.
        - System analyzes how current topics relate to other topics searched by analyst/user in past (information of interest to analyst/user).

        & - The system provides an extractive summary as well as most influential content in the output window. But the system will not provide any further information on the content used in the summary, such as details about the user accounts, location of content creator, or timestamp of content.
      
        - The system cannot identify trends for follow up work.\\
      
        \midrule 
        \textbf{Logic}   
        & - System can interpret data correctly and produce conclusions that are backed up by data.
      
        - System can demonstrate how it derived its results, summary is clearly based on cited evidence.
      
        - System provides information on logic used to derive the results and shows the source documents used to derive summary (covered in Sourcing criteria).
      
        - Logical approach, consistent and meaningful.
      
        - System uses appropriate language for intended analysts/users, displays results in a logical progression and results are contextualized in an understandable way.
      
        - System displays topics derived from analysis and allows user to drill down to see underlying titles/summaries, reporting used to derive titles/summaries and other supplementary information (covered in Sourcing criteria).
      
        - System provides detailed description of contrary or outlying data and why it chose the given result over other possibilities (covered in Uncertainty and Analysis of Alternatives criteria).
      
        - System applies notable skill or sophistication in deriving results by combing different data, contexts, assumptions.
      
        - System describes methodological framework from which it derives judgments and the evidence and assumptions used are relevant to the results.

        & - The system does not display results in a logical progression and results are not contextualized in a way the user can understand. All content is mixed up in the summary window and it is unclear, for example, if a post title is actually a tweet. \\
        
        \midrule 
        \textbf{Change}  
        & - System notes changes or consistency with previous results.
        
        - System provides a ‘similarity score’ between current and previous summaries on similar searches and notes any semantic shifts.
        
        - System explains in simple terms how the new data caused system to maintain or change past derived results.
        
        - System compares current results with results from outside systems or agencies on similar topics and describes similarities and differences to user/analyst.
        
        & - The system does not provide a way to create additional summaries and compare them with each other. The system also does not provide a way to note any changes with previous analyses or similar topics.   \\
        
        \midrule 
        \textbf{Accuracy}    
        
        & - The system’s derived results are clear and understandable.
        
        - The system’s derived result factors in timing, likelihood and other conditional factors supporting an actionable conclusion.
        
        - There are no ill-defined conditions and analyst/user is comfortable that if there were conditions, they would be expressed by the system.
                
        - The system presents a measure of accuracy (percent confidence that a judgment is correct) which is sufficiently high.
        
        - Analyst/user can independently determine that the systems results are accurate, complete, and consistent.
        
        - Analyst/user can select portions of summary and test for accuracy.
        
        & - The system does not display results in a logical progression and rather presents them in a way that is confusing to the user. Different types of content are mixed together in a disorganized fashion. The summary displays mostly content related to the keywords used, but some content is remotely related. This makes the resultant summary non-actionable for a user, since it is not possible to simply remove irrelevant content because it is not possible to determine why that content was included (i.e. the criteria for being included in the summary).\\

        \midrule 

        \textbf{Visualization}   
        & - Includes understandable and directable visual information where appropriate.
        
        - Visual is pertinent to information conveyed in summary and helps show key concepts (like graphs, scatterplots, or visual that links items grouped in a topic).
        
        - Visual has proper labels and data points are explained.
        
        - Analyst/user can learn something new from visual produced.
        
        - Visualization highlights data to clarify, complement, or augment understanding.
        
        - The system chooses the most effective type of visual to display because it knows the analysts’ goals and data available.
        
        - The system can display micro and macro trends in visuals in a way that is easily digestible and with a visual representation of why trends are significant.
        
        - The visual displayed does not require analyst/user to look through other sources, etc. to understand and it does not overly reference other addendums, etc. 
        & - The system cannot provide any visualizations.\\
    \bottomrule 
    \caption{The goals of High-MAST and Low-MAST READIT 1.0. }
    \label{tab:tableAppA} \\
\end{longtable}

%% file: sections/appendixB.tex
\newpage 
\section{The Nine MAST Criteria for READIT 2.0}
\label{appendixB}


\begin{longtable}{p{0.2\linewidth}|p{0.7\linewidth}}

    \toprule
    \textbf{MAST item} & \textbf{Questions and Feature Descriptions} \\
    \midrule
    
    
    \multirow{3}{*}{\textbf{Sourcing}} & How well can the system identify underlying sources and methodologies upon which results are based?  \\
        & \textbf{High-MAST:} In the \textit{Documents} page, you can see descriptive information about the documents (data) used to gather the clusters including basic information and detailed descriptions of the sources. The \textit{About} page includes information on the clustering model, models for summarization, training data, possible biases, pre-processing of data, and quality of the data used in training to derive results. In the main dashboard view, you can view the data used to derive the cluster either by hovering or clicking on it including the cluster title, number of documents, top terms, and representative documents. The representative documents can be viewed as a summary (derived result) or raw version.\\ 
        & \textbf{Low-MAST:} For any given cluster in the main dashboard view, you can view more details about it by clicking on it. The title of the cluster, number of documents, and summaries of the documents will be displayed in the documents and summaries pane. Only the derived results are shown, not the underlying sources and data used to derive the clusters or summaries.\\
    
      \midrule 
      \multirow{3}{*}{\textbf{Uncertainty}} & How well can the system indicate and explain the basis for the uncertainties associated with derived results? \\
        & \textbf{High-MAST:} READIT indicates levels of uncertainty with derived results in two ways, as described in the \textit{About} page. First, READIT includes keywords per cluster to show how documents in clusters are related to each other. Keywords are displayed with a term frequency–inverse document frequency (tf-idf) score which measures the certainty the word fits with the cluster. Second, READIT includes similarity scores to assess the similarity between clusters. This score is calculated using cosine similarity to show the certainty that clusters are related to each other.\\
        & \textbf{Low-MAST:} In the topic similarity visualization, the relationship between two topics is colored from white to dark blue with dark blue indicating a higher certainty the two topics are related. These relationships are not labeled with numbers, neither is it explained how this similarity is calculated.\\

      \midrule 
      \multirow{3}{*}{\textbf{Distinguishing}}  & How well can the system clearly distinguish derived results and underlying data?\\
        & \textbf{High-MAST:} For any given cluster you can view more details about the data used to derive the cluster either by hovering or clicking on it. The \textit{About} page for READIT includes information on the clustering model, models for summarization, training data, underlying assumptions for choice of training data, quality of the data used in training to derive results, possible biases, pre-processing of data, recommended uses and users, and restrictions on use. The \textit{About} page was created with domain expert input.\\
        & \textbf{Low-MAST:} When opening or clicking on clusters, you can view more details about that cluster. The title and summary of representative documents will appear. The raw data used to derive the title and summaries is not displayed. There is no \textit{About} page with information on how these titles or summaries are calculated. \\
      
      \midrule 
      \multirow{3}{*}{\textbf{\makecell{Analysis of \\Alternatives}}}    & How well can the system identify and assess plausible alternative results?\\
        & \textbf{High-MAST:} In the topic similarity visualization, users initially view the visualization where the topics are ordered alphabetically. By factoring in the similarity score and uncertainties, READIT can reorder the view in this visualization such that highly related topics will appear together to present an alternative view.\\
        & \textbf{Low-MAST:} READIT is not able to show alternative results when uncertainties in the data warrant them. There is no way to reorder visualizations based on any criteria.\\
      
      \midrule 
      \multirow{3}{*}{\textbf{\makecell{Customer \\Relevance}}}  & How well can the system provide information and insight to users?\\
        & \textbf{High-MAST:} READIT synthesizes large corpora of documents and produces clusters of similar documents. The topic similarity visualization shows which clusters are most highly related to each other. Users can examine the clusters and their relationships in the topic similarity view for trends for follow-up work. READIT is also able to suggest locations to filter by if the documents contain multiple locations. Users can also filter all visualizations by topic. There is a topic filtering pane where users can check all, or some topics and the corresponding selected topics will be highlighted in the visualizations.\\
        & \textbf{Low-MAST:} READIT synthesizes large corpora of documents and produces clusters of similar documents. The topic similarity visualization shows which clusters are most highly related to each other. Users can examine the clusters and their relationships in the topic similarity view for trends for follow-up work.\\
      
      \midrule 
      \multirow{3}{*}{\textbf{Logic}}   & How well can the system help the user understand how it derived its results?\\
        & \textbf{High-MAST:} For any given cluster you can view more details about the data used to derive the cluster either by hovering or clicking on it. The \textit{About} page includes information on pre-processing of data. READIT includes an option to filter results by location if location information is detected in the document. To give location options, READIT must consider the location information in the context of the document, and other assumptions about the embedding of the location in the document.\\
        & \textbf{Low-MAST:} When clicking on clusters in the main view, you can view the title and representative documents in summary form. The titles and summaries are understandable to users. Information on how clusters, titles, and summaries are formed is not included. There is also no information on the pre-processing of data.\\    
      
      \midrule 
      \multirow{3}{*}{\textbf{Change}}  & How well can the system help the user understand how derived results on a topic are consistent with or represent a change from previous analysis of the same or similar topic?\\
        & \textbf{High-MAST:} In the \textit{Documents} page, READIT includes information on similar searches from other agencies. Similar searches may be based on the average length of the document, number of documents, or number of clusters generated.\\
        & \textbf{Low-MAST:} READIT does not have a way to note changes from previous analyses or similar analyses. READIT also cannot compare current results with those of other agencies which had similar results.\\
      
      \midrule 
      \multirow{3}{*}{\textbf{Accuracy}}    & How well can the system make the most accurate judgments and assessments possible, based on the information available and known information gaps?\\
        & \textbf{High-MAST:} The READIT \textit{About} page includes information on system verification and validation methodology, and results from the training data where the system achieved sufficiently high accuracy. To assess the accuracy of READIT, users can view the full documents used in each cluster and compare them against the top terms to independently determine whether the documents match the top terms. Likewise, users can view a summary of the document and compare it against the full version of the document in the documents and summaries view to see if the summary is accurate.\\
        & \textbf{Low-MAST:} READIT does not include information on system verification, validation methodology, or information on the training of the system where it achieved sufficient accuracy. Since underlying sourcing information and raw data are not included in the system, it is difficult to assess whether the topics and summaries are accurate.\\
      
      \midrule 
      \multirow{3}{*}{\textbf{Visualization}}   & How well can the system incorporate visual information if it will clarify an analytic message and complement or enhance the presentation of data and analysis? Is visual information clear and pertinent to the product’s subject?\\
        & \textbf{High-MAST:} READIT uses three main visualizations to enhance users’ understanding of the clusters. First, in the topic overview visualization, clusters are displayed as bubbles where the size of the bubbles can indicate anomalies. Next, READIT also creates and displays a topic similarity visualization to help understand the connections between clusters. Lastly, there is a timeline view in READIT to display clusters on a timeline (if documents contain date information). All visualizations are simple and labeled properly. Users can view more details about the visualizations by clicking on them or hovering over them or filtering all visualizations by cluster using the filtering option.\\
        & \textbf{Low-MAST:} READIT uses two visualizations. The similarity matrix shows the similarity scores between topics. Darker colors indicate more similarity but score values are not shown. The timeline shows the clusters on the timeline. Visualizations contain no interactivity and users are not able to click or hover on items to view more details about the visualizations. \\
    \bottomrule 
    \caption{MAST and READIT 2.0 feature descriptions for High-MAST and Low-MAST}
    \label{tab:tableAppB} \\
\end{longtable}

%% file: sections/appendixC.tex
\newpage 
\section{Study Questionnaires}
\label{appendixC}


\begin{longtable}{p{0.25\linewidth}|p{0.34\linewidth}|p{0.2\linewidth}|p{0.1\linewidth}}
    \toprule
    \textbf{Variables} & \textbf{Example item(s)} & \textbf{Number of items  / Reverse items} & \textbf{Scale} \\
    \midrule
    \makecell[l]{\textbf{MAST-total}\\ \citep{blasch2021multisource}} & Sourcing, uncertainty, distinguishing, analysis of alternatives, customer relevance, logical argumentation, consistency, accuracy, and visualization & 9/0 & 9 - 36 \\
    
    \midrule 
    \makecell[l]{\textbf{Trust (Jian)}\\ \citep{jian2000foundations}} & “I can trust the system.”; “The system looks deceptive.” & 12/5 & 1 - 7 \\
    
    \midrule 
    \makecell[l]{\textbf{Trust (Chancey)}\\ \citep{chancey2017trust}} & “I understand how the system will help me perform well. “; “The information the system provides reliably helps me perform well. & 15/0 & 1 - 7 \\

    \midrule 
    \makecell[l]{\textbf{Engagement}\\ \citep{schaufeli2002measurement}} & “I was immersed in this research task.”; “To me, this research task was challenging.” & 17/0 & 1 - 7 \\
    
    \midrule 
    \makecell[l]{\textbf{Usability (SUS)}\\ \citep{brooke2020sus}} & “I felt very confident using the system.”; “I thought the system was easy to use.” & 10/5 & 1 - 5 \\
      \bottomrule 
      
  \caption{Dependent and Control Variables}
  \label{tab:tableB1} \\
\end{longtable}

%% file: sections/appendixE.tex
\newpage
\section{Data and Matrices Used in Principal Components Analysis}
\label{appendixE}

\begin{table*}[h!]
\newcommand{\StyledMatrix}[3]{
\left[
\begin{array}{c|*{#2}{c}}
    & #3 \\
    \hline
    #1
\end{array}
\right]
}

\newcommand{\HeaderRowA}{\scriptstyle S_3 & \scriptstyle S_4 & \cdots & \scriptstyle V_7}
\newcommand{\HeaderRowB}{\scriptstyle S_2 & \scriptstyle U_3 & \cdots & \scriptstyle V_5}
\newcommand{\HeaderRowC}{\scriptstyle L_1 & \scriptstyle L_4 & \cdots & \scriptstyle U_2}
\newcommand{\HeaderRowM}{\scriptstyle S_1 & \scriptstyle S_2 & \cdots & \scriptstyle V_7}

\newcommand{\BodyA}{
    \scriptstyle u_1 & 2 & 2 & \dots & 2 \\
    \scriptstyle u_2 & 3 & 1 & \dots & 4 \\
    \vdots & \vdots & \vdots & \ddots & \vdots \\
    \scriptstyle u_m & 2 &  & \dots & \\
}

\newcommand{\BodyB}{
    \scriptstyle u_1 & 5 & 3 & \dots & 1 \\
    \scriptstyle u_2 & 2 & 4 & \dots & 3 \\
    \vdots & \vdots & \vdots & \ddots & \vdots \\
    \scriptstyle u_m & 1 &  & \dots & \\
}

\newcommand{\BodyC}{
    \scriptstyle u_1 & 3 & 1 & \dots & 4 \\
    \scriptstyle u_2 & 2 & 2 & \dots & 2 \\
    \vdots & \vdots & \vdots & \ddots & \vdots \\
    \scriptstyle u_m & 3 & 3 & \dots & 3 \\
}

\newcommand{\BodyM}{
    \scriptstyle u_1 & \cdot & \cdot & \dots & \cdot \\
    \scriptstyle u_2 & \cdot & \cdot & \dots & \cdot \\
    \vdots & \vdots & \vdots & \ddots & \vdots \\
    \scriptstyle u_m & \cdot & \cdot & \dots & \cdot \\
}

\begin{center}
\begin{tikzpicture}
\def\shiftX{0.50}
\def\shiftY{-0.8}

\node (A) at (0,0) {$
    \StyledMatrix{\BodyA}{4}{\HeaderRowA}
$};
\node (LabelA) at ($(A.east)+(2.0cm,0)$) {Performance ($\boldsymbol{\mathrm{P_1}})_{m\times n_1}$};

\node (B) at ($(A.south east)+(-\shiftX,\shiftY)$) {$
    \StyledMatrix{\BodyB}{4}{\HeaderRowB}
$};
\node (LabelB) at ($(B.east)+(1.50cm,0)$) {Process ($\boldsymbol{\mathrm{P_2}})_{m\times n_2}$};

\node (C) at ($(B.south east)+(-\shiftX,\shiftY)$) {$
    \StyledMatrix{\BodyC}{4}{\HeaderRowC}
$};
\node (LabelC) at ($(C.east)+(1.5cm,0)$) {Purpose ($\boldsymbol{\mathrm{P_3}})_{m\times n_3}$};

\node[anchor=east] (M) at ($(B.west)+(-1cm,0)$) {
    \raisebox{-10pt}{\large MAST Ratings}
    $\underset{\boldsymbol{\mathrm{U}}_{m\times n}}{
        \StyledMatrix{\BodyM}{4}{\HeaderRowM}
    }$
};

\draw[->, thick] (M.east) -- (B.west);

\node (DoubleArrow) at ($(C.south)+(-2,-1.5cm)$) {
    \begin{tikzpicture}
        \draw[double, ->, thick] (0,0) -- (-1,-1);
    \end{tikzpicture}
};

\end{tikzpicture}
\end{center}

\begin{center}
\[
\begin{array}{ccc}
\underset{\raisebox{0pt}{\text{\large $\boldsymbol{(\mathrm{P_i})_{m\times n_i}}$}}}{
\left[
\begin{array}{c|cccc}
    & \scriptstyle S_3 & \scriptstyle S_4 & \cdots & \scriptstyle V_7 \\
    \hline
    \scriptstyle u_1 & 2 & 2 & \dots & 2 \\
    \scriptstyle u_2 & 3 & 1 & \dots & 4 \\
    \vdots & \vdots & \vdots & \ddots & \vdots \\
    \scriptstyle u_m & 2 & 1 & \dots & 2 \\
\end{array}
\right]
}
& \times &
\underset{\raisebox{0pt}{\text{\large $\boldsymbol{(\mathrm{F})_{n_i\times p}}$}}}{
\left[
\begin{array}{c|cccc}
    & \scriptstyle Doc & \scriptstyle Abt & \cdots & \scriptstyle Time \\
    \hline
    \scriptstyle S_3 & 1 & 1 & \dots & 1 \\
    \scriptstyle S_4 & 0 & 0 & \dots & 0 \\
    \vdots & \vdots & \vdots & \ddots & \vdots \\
    \scriptstyle V_7 & 0 & 0 & \dots & 1 \\
\end{array}
\right]
}
\\[2ex]
\multicolumn{3}{c}{
=
}
\\[2ex]
\multicolumn{3}{c}{
\underset{\raisebox{0pt}{\text{\large $\boldsymbol{(\mathrm{T_i})_{m\times p}}$}}}{
\left[
\begin{array}{c|cccc}
    & \scriptstyle Doc & \scriptstyle Abt & \cdots & \scriptstyle Time \\
    \hline
    \scriptstyle u_1 & 2 & 4 & \dots & 8 \\
    \scriptstyle u_2 & 3 & 2 & \dots & 10 \\
    \vdots & \vdots & \vdots & \ddots & \vdots \\
    \scriptstyle u_m & 2 & 2 & \dots & 8 \\
\end{array}
\right]
}
}
\end{array}
\]
\end{center}

\caption{The schematic shows how the participants' MAST ratings were mapped to system-level features and trust constructs (process, performance, purpose).}
\label{tab:matrix_mult}
\end{table*}

%% file: sections/appendixD.tex
\newpage 
\section{PCA Evaluation Results}
\label{appendixD}

\begin{figure*}[!h]
    \centering
    \begin{subfigure}[b]{0.4\textwidth}
        \centering
        \includegraphics[width=\textwidth]{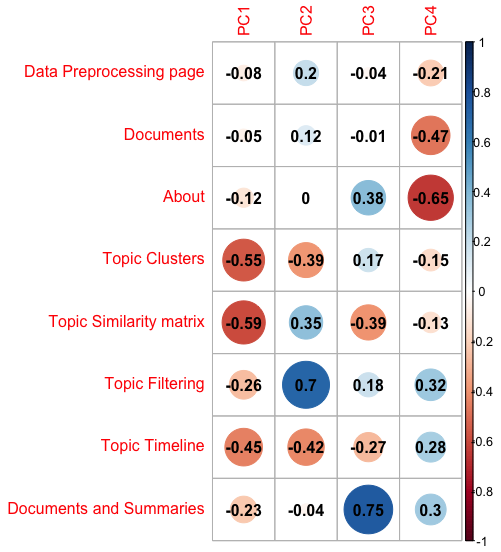}
        \caption{Performance}
    \end{subfigure} \hfill
    \begin{subfigure}[b]{0.4\textwidth}
         \includegraphics[width=\textwidth]{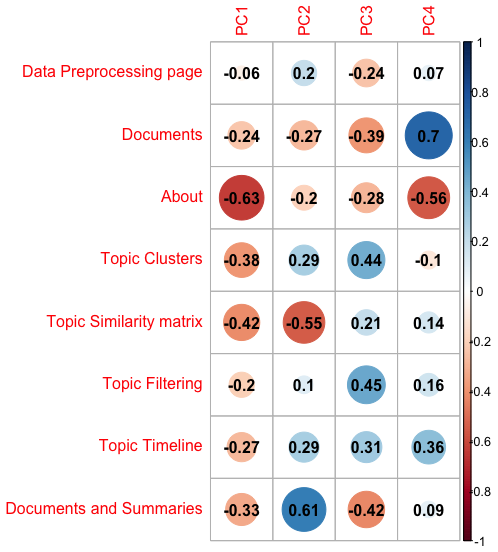}
         \caption{Process}
    \end{subfigure}
    \begin{subfigure}[b]{0.4\textwidth}
         \includegraphics[width=\textwidth]{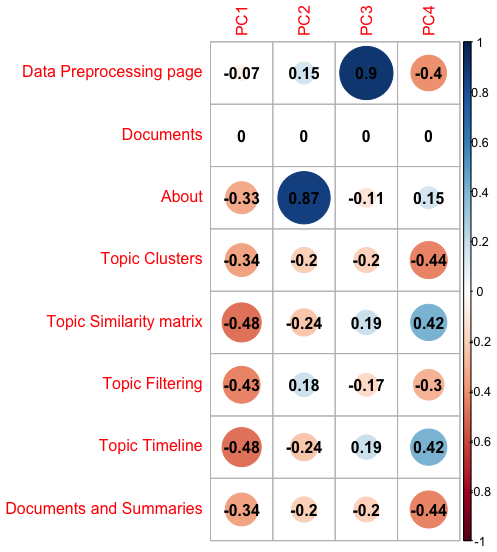}
         \caption{Purpose}
    \end{subfigure}
    \caption{PCA loadings for three dimensions in trust theory}
    \label{fig:pca}
\end{figure*}

\begin{figure*}[t]
    \centering
    \begin{subfigure}[b]{0.45\textwidth}
        \centering
        \includegraphics[width=\textwidth]{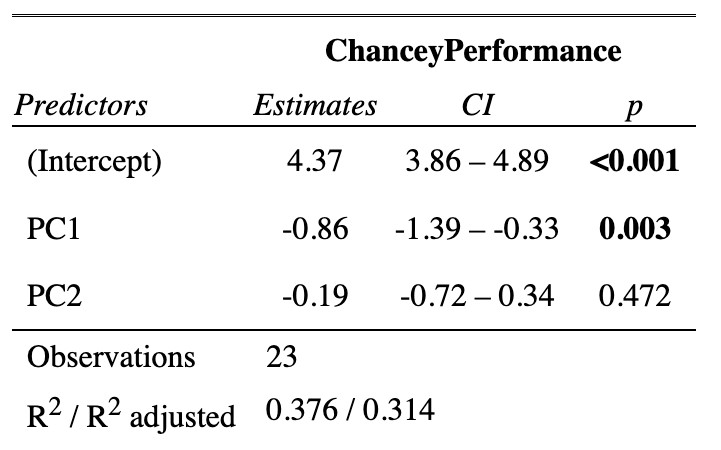}
        \caption{Performance}
    \end{subfigure} \hfill
    \begin{subfigure}[b]{0.45\textwidth}
         \includegraphics[width=\textwidth]{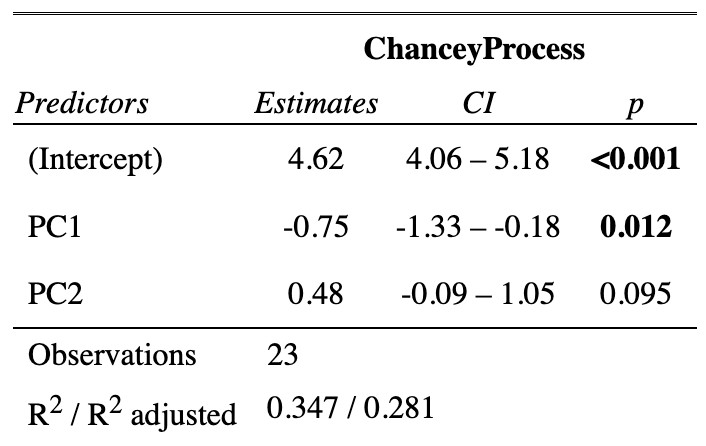}
         \caption{Process}
    \end{subfigure} 
    
    \begin{subfigure}[b]{0.45\textwidth}
         \includegraphics[width=\textwidth]{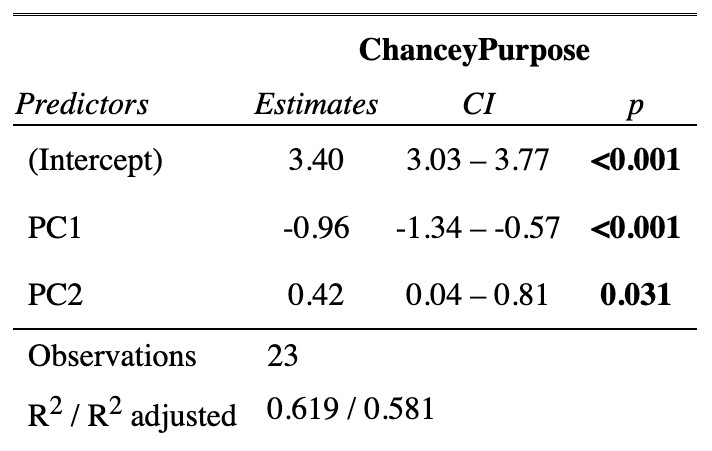}
         \caption{Purpose}
    \end{subfigure}
    \caption{Principal Component Regression results}
    \label{fig:lm}
\end{figure*}





